\documentstyle[preprint,prc,aps]{revtex}
\newcommand{\bsigma}{\mbox{\boldmath $\sigma$}}
\newcommand{\btau}{\mbox{\boldmath $\tau$}}

\begin{document}

\draft

\title{Ground state correlations in $^{16}$O and $^{40}$Ca}
\author{A. Fabrocini$^{1)}$, F.Arias de Saavedra$^{2)}$ and G.Co'$^{\,3)}$}  
\address{
$^1)$ Dipartimento di Fisica, Universit\`a di Pisa,\\ 
and Istituto Nazionale di Fisica Nucleare, sezione di Pisa,\\
I-56100 Pisa, Italy \protect\\
$^2)$ Departamento de Fisica Moderna, 
Universidad de Granada, \\
E-18071 Granada, Spain \protect\\
$^3)$
Dipartimento di Fisica, Universit\`a di Lecce \\
 and Istituto Nazionale di
Fisica Nucleare, sezione di Lecce,  \\  
I-73100 Lecce, Italy }
\maketitle

\date{\today}

\begin{abstract}

We study the ground state properties of doubly closed shell 
nuclei $^{16}$O and $^{40}$Ca in the framework of 
Correlated Basis Function theory using state dependent correlations, 
with central and tensor components. 
The realistic Argonne $v_{14}$ and $v'_{8}$ two--nucleon potentials 
and three-nucleon potentials of the Urbana class have been adopted. 
By means of the Fermi Hypernetted Chain integral equations, 
in conjunction with the Single Operator Chain approximation, 
we evaluate the ground state energy, one-- and two--body densities  
and electromagnetic and spin static 
responses for both nuclei. In $^{16}$O we compare our results 
with the available Monte Carlo and Coupled Cluster ones and find a 
satisfying agreement. As in the nuclear matter case with similar 
interactions and wave functions, the nuclei result under-bound by 
2--3 MeV/A.

\end{abstract}


%
%
\narrowtext

\section{Introduction}

The attempt to describe all nuclei
starting from the same nucleon--nucleon interaction which reproduces the
properties of two--, and possibly three--, nucleon systems is 
slowly obtaining its first successes. 
A set of techniques to exactly solve the Schr\"odinger equation 
in the 3$\leq$A$\leq$8 nuclei is now available:
Faddeev \cite{che86}, Correlated Hyperspherical Harmonics Expansion
\cite{kie93}, Quantum Monte Carlo \cite{pud97}. 
Their straightforward extension to medium-heavy nuclei 
is however not yet feasible, both for computational and theoretical 
 reasons. 

The Correlated Basis Function (CBF) theory is one of the most 
promising many--body tools currently under development to 
attack the problem  of dealing with the complicate structure 
(short range repulsion and strong state dependence) of the nuclear 
interaction. The CBF has a long record of applications in 
condensed matter physics, 
as well as in liquid helium and electron systems. 
In nuclear physics the most extensive use of CBF has been done in  
infinite nuclear and neutron matter. 
The neutron stars structure described via the CBF based neutron matter 
equation of state is in nice agreement with the    
current observational data \cite{wir88,akm98}. 
In nuclear matter CBF has been used not only to study ground state
properties \cite{wir88,fri81,akm97}  but also dynamical quantities, as 
electromagnetic responses \cite{fab89,fab97} and one-body Green's 
functions \cite{ben92}.

The CBF theory is based upon the variational principle, i.e. one
searches for the minimum of the energy functional
\begin{equation}
\label{enfun}
E[\Psi]=\frac {\langle \Psi|H|\Psi\rangle}{\langle \Psi|\Psi\rangle} 
\end{equation}
in the Hilbert subspace of the correlated many--body wave functions $\Psi$: 
\begin{equation}
\label{ansatz}
\Psi(1,2...A)= G(1,2...A)\Phi(1,2...A),
\label{Psi}
\end{equation}
where  $G(1,2...A)$ is a many--body correlation operator acting on the 
mean field wave function $\Phi(1,2...A)$ (we will take a Slater 
determinant of  single particle wave functions, $\phi_\alpha(i)$). 
In realistic 
nuclear matter calculations, the correlation operator is given by 
a symmetrized product of two-body correlation operators, $F_{ij}$, 
\begin{equation}
G(1,2...A)={\cal S}\left[\prod_{i<j}F_{ij}\right].
\label{G}
\end{equation}
In principle richer choices for the operator (\ref{G}) can be made  
 by including explicit three--, or more-- , nucleon correlations, 
which cannot be described by the product of two--body correlations.
It is essential, however, that the  two--body correlation $F_{ij}$
has an operatorial dependence analogous to that of the 
modern nucleon--nucleon interactions
\cite{arn92,sto93,sto94,wir95,mac96}.
Nowadays, sophisticated CBF calculations 
consider $F_{ij}$ of the form: 
\begin{equation}
\label{corr8}
F_{ij}=\sum_{p=1,8}f^p(r_{ij})O^p_{ij},
\end{equation}
where the involved operators are:
\begin{equation}
\label{oper8}
O^{p=1,8}_{ij}=
\left[ 1, \bsigma_i \cdot \bsigma_j, S_{ij}, 
({\bf L} \cdot {\bf S})_{ij} \right]\otimes
\left[ 1, \btau_i \cdot \btau_j \right] 
\end{equation}
and $S_{ij}=(3\,{\hat  {\bf r} }_{ij} \cdot \bsigma_i  \,
{\hat{\bf r}}_{ij} \cdot 
\bsigma_j -  \bsigma_i \cdot \bsigma_j)$ is the tensor operator. 
The correlation functions $f^p(r)$, as well as the
set of single particle wave functions, are fixed by the energy minimization 
procedure.

A key point in applying CBF is the evaluation of the many--variables
integrals necessary to calculate the energy functional (\ref{enfun}).
A direct approach consists in using Monte Carlo sampling 
techniques (Variational MonteCarlo,  VMC) \cite{wir91}. 
However, the required numerical effort is such that,
for realistic interactions and correlations, VMC
can be efficiently used only in light nuclei. Actually, 
a realistic calculation of the ground state of $^{16}$O
has been done in Ref.\cite{pie92} by using the so called 
Cluster MonteCarlo (CMC) method. In CMC 
the terms related to the scalar part of the correlation ($p=1$) 
are completely summed by VMC, whereas the remaining 
operatorial ($p>1$) contributions are approximated by 
considering up to four-- or five--body cluster terms. 

An alternative to the MonteCarlo methodology is provided by cluster 
expansions and the integral summation technique known as Fermi HyperNetted 
Chain (FHNC) \cite{ros82}, particularly suited to treat heavy systems.
By means of the FHNC equations it is possible to 
sum infinite classes of Mayer-like diagrams resulting from the cluster 
expansion of the expectation value of the hamiltonian, 
or of any other operator. FHNC has been  widely applied to
both finite and infinite systems with purely scalar
(state independent, or Jastrow) correlations.  

The case of the state dependent $F_{ij}$, needed in  
nuclear systems, is more troublesome since the non commutativity of the 
correlation operators prevents from the development of a complete 
FHNC theory for the correlated wave function of Eq.(\ref{Psi}).
For this reason an approximated treatment of the operatorial 
correlations, called Single Operator Chain (SOC), has been developed
\cite{pan79}.  The SOC approximation, together with
a full FHNC treatment of the Jastrow part of the correlation, provides 
an accurate description of infinite nucleonic matter \cite{wir88}. It
is therefore believed that FHNC/SOC effectively includes the
contribution of many--body correlated clusters at all orders.
The evaluation of additional classes of 
diagrams in nuclear matter has set the estimated accuracy for 
the ground state energy to less than $1$ MeV at saturation density 
($\rho_{nm}=$0.16 fm$^{-3}$) \cite{wir88,wir80}.

In a series of papers \cite{co92,co94,ari96}
we have extended the FHNC scheme to describe 
the ground state of doubly closed shell nuclei, from $^4$He to $^{208}$Pb, 
with semi-realistic, central interactions and two-body correlations, 
either of the Jastrow type or depending, at most, on the third 
components of the isospins of the correlated nucleons. 
In Ref.\cite{fab98} we used FHNC/SOC 
to evaluate energies and densities of the $^{16}$O and $^{40}$Ca
nuclei, having doubly closed shells in the {\em ls} coupling scheme,
with potentials and correlations containing operator terms 
up to the tensor components. In the $^{16}$O nucleus the
comparison of our results with those of a CMC calculation
confirmed the accuracy of the FHNC/SOC approximation estimated in
nuclear matter. 

The present work is the extension of that of Ref. \cite{fab98}.
The ground state properties of the $^{16}$O and $^{40}$Ca nuclei are
calculated within the FHNC/SOC formalism by using a complete, 
realistic nucleon--nucleon potential, with $p>6$ components,  
and by considering also three--nucleon interactions.
The two--nucleon interactions we have employed are
the Argonne $v_{14}$ \cite{wir84} potential  and the  $v'_{8}$ 
reduction of the Argonne $v_{18}$ \cite{wir95} potential. 
 For the three--nucleon interaction we have 
adopted the Urbana models,  Urbana VII \cite{sch86} with 
Argonne $v_{14}$  and  Urbana IX
\cite{pud97} with  Argonne $v'_{8}$.   
In addition to the energy and the densities, we have also evaluated 
the static responses. They  are 
the non energy weighted sums of the inclusive dynamical responses of 
the nucleus to external probes. We have studied the density, the 
electromagnetic and the spin static responses, both in the longitudinal 
and transverse channels.

The paper is organized as follows: 
in section 2 we briefly present
the interaction and the correlated wave function properties  
and recall the basic features of FHNC/SOC; section 3 deals in 
short with the insertion of the spin--orbit components and of the 
three--nucleon potential; in section 4 we show and discuss
the results for the energy, one-- and two--body densities  
and static responses; the conclusions are drawn in section 5.

\section{Interaction, correlated wave function and cluster expansion}

We work in the framework of the non relativistic 
description of the nucleus and use a hamiltonian of the form:
\begin{equation}
H={{-\hbar ^2}\over2\,m}\sum_i\nabla_i^2+\sum_{i<j}v_{ij}
+\sum_{i<j<k}v_{ijk} \,\, .
\label{hamilt}
\end{equation}
Very high quality phase--shift analyses of a large body of
$pp$ and $np$ data have been recently carried out \cite{arn92,sto93}.
 Building on this accurate data base, several nucleon--nucleon (NN)
 potentials have been constructed, 
like the updated Nijmegen interaction \cite{sto94},
the CD Bonn interaction \cite{mac96}
and the Argonne $v_{18}$ (A18) \cite{wir95} interaction. All of them include
charge symmetry breaking terms in order to provide a precise fit
to both the $pp$ and $np$ data. 

The structure of $v_{ij}$ and $v_{ijk}$, at large inter-particle distances 
is dictated by meson exchange processes. The long--range part of the NN 
interactions is determined by the one--pion exchange (OPE): 
\begin{equation}
v^{OPE}_{ij}={{f^2_{\pi NN}}\over4\pi}{{m_\pi}\over3}
\left[ Y_\pi(r_{ij})\bsigma_i \cdot \bsigma_j +
       T_\pi(r_{ij})S_{ij}\right]\btau_i \cdot \btau_j \,\,.
\label{opep}
\end{equation}
where $m_\pi$ is the pion mass (138.03 MeV), 
$(f^2_{\pi NN}/4\pi)=0.081$, and
$Y_\pi (r)$ and $T_\pi (r)$ are the Yukawa and tensor Yukawa functions,  
\begin{equation}
Y_\pi (r)={{e^{-\mu r}}\over{\mu r}} ,
\label{ypi}
\end{equation}
\begin{equation}
T_\pi (r)={{e^{-\mu r}}\over{\mu r}}
\left[ 1 + {3\over{\mu r}} + {3\over{(\mu r)^2}} \right] ,
\label{tpi}
\end{equation}
with  $\mu\sim 0.7$ fm$^{-1}$.
The Argonne potentials simulate the
effects of $\rho$ exchange, not explicitly included 
\cite{wir84}, by modifying the Yukawa functions with 
a short--range cutoff,
$Y_\pi(r)\rightarrow Y(r)=Y_\pi(r) F_{cut}(r)$ and
$T_\pi(r)\rightarrow T(r)=T_\pi(r) F^2_{cut}(r)$ with
$F_{cut}(r)=1-\exp (-cr^2)$ ($c$=2 fm$^{-2}$). 
The intermediate and short--range 
parts of this class of potentials are mostly phenomenological and 
A18 is parametrized according to the operatorial 
structure:
\begin{equation}
v_{ij}=\sum_{p=1,18}v^p(r_{ij})O^p_{ij},
\label{argo18}
\end{equation}
where the first 14 components, 
\begin{equation}
O^{p=1,14}_{ij}=
\left[ 1, \bsigma_i \cdot \bsigma_j, S_{ij}, 
({\bf L} \cdot {\bf S})_{ij}, L^2, L^2 \bsigma_i \cdot \bsigma_j,
({\bf L} \cdot {\bf S})^2_{ij} \right]\otimes
\left[ 1, \btau_i \cdot \btau_j \right] ,
\label{op14}
\end{equation}
 give the  isoscalar part, defining a $v_{14}$--like potential (A14),
\begin{equation}
v_{14,ij}=\sum_{p=1,14}v^p(r_{ij})O^p_{ij}.
\label{v14}
\end{equation}
The remaining 4 components of A18 are of the isovector
($\tau_{i,z}+\tau_{j,z}$) and isotensor
($3\tau_{i,z}\tau_{j,z}-{\bf\tau}_i\cdot{\bf\tau}_j$) type.
The Argonne $v'_{8}$ potential (A8') is an eight operators 
reduction of A18 built to reproduce the isoscalar part of the full
interaction  in the $S$, $P$ and $^3D_1$ waves and the $^3D_1-^3S_1$ 
coupling \cite{pud97}. 

Other potentials  use different parameterizations.  
For instance, the Nijmegen model \cite{sto93} employs 
${\bf p}^2$ operators instead  of ${\bf L}^2$.  

The longest range part of the three--nucleon interactions (TNI) involves 
a two pions exchange with the intermediate excitation of a $\Delta$ 
\cite{fuj57}: 
\begin{equation}
v^{2\pi}_{ijk}=A_{2\pi} \sum_{cyc}\left(
\{ X_{ij}, X_{ik} \} 
\{ \btau_i \cdot \btau_j,\btau_i \cdot \btau_k\}
+ {1\over 4}
\left[ X_{ij}, X_{ik} \right] 
\left[ \btau_i \cdot \btau_j,\btau_i \cdot \btau_k\right]
\right) ,
\label{tpi2p}
\end{equation}
where
\begin{equation}
X_{ij}=Y(r_{ij})\bsigma_i \cdot \bsigma_j +
       T(r_{ij})S_{ij} \,\, ,
\label{xij}
\end{equation}
and the symbols $[,]$ and $\{,\}$ indicate the commutator and
anti--commutator operators respectively.

The Urbana class of TNI \cite{car83} introduces an additional, repulsive, 
spin and isospin independent, short range term,
\begin{equation}
v^{R}_{ijk}=U_0 \sum_{cyc} T^2(r_{ij})T^2(r_{ik}) ,
\label{tpir}
\end{equation}
which simulates dispersive effects when integrating out $\Delta$
degrees of freedom. The total Urbana TNI are then given by the sum of
the two terms defined above, 
$v_{ijk}= v^{2\pi}_{ijk} + v^{R}_{ijk}$, 
and the $A_{2\pi}$ and $U_0$ parameters are adjusted to provide a good fit 
to the binding energies of few--body nuclei and nuclear matter. In the 
A14+Urbana VII (A14+UVII) model the values $A_{2\pi}$= -0.0333 MeV and 
$U_0$= 0.0038 MeV have been fixed by variational calculations. In the 
most recent A18+Urbana IX (A18+UIX) model the values 
$A_{2\pi}$= -0.0293 MeV and $U_0$= 0.0048 MeV have been obtained with 
a Quantum MonteCarlo calculation for $^3$H and a variational calculation 
for nuclear matter. 
The values of  $A_{2\pi}$ are in good 
agreement with those predicted by the pure two--pions exchange model 
($A_{2\pi}\sim$ -0.03 MeV). 

The truncated A8' NN potential was introduced in Ref. \cite{pud97} 
because its simpler parameterization allowed a simplification of 
the numerically involved Quantum MonteCarlo calculations. 
The contribution of the missing channels was  perturbatively
evaluated. One should remark, however, that A8' was found to
give a slight over bind. 
For this reason, the strength of the repulsive part of the 
TNI Urbana IX model was increased by $\sim$30$\%$ to reproduce the 
experimental energies.
The results presented in this paper have been obtained with the
A14+UVII and A8'+UIX models, where the Urbana IX interaction has been
redefined as above.

The many--body wave function (\ref{Psi}) contains two ingredients:  
the correlation functions $f^p(r)$ and the single particle states
forming the Slater determinant $\Phi(1,2...A)$. 
In our calculations we have used a $f_6$--type correlation, so 
neglecting the spin--orbit components.
This choice will be discussed and commented later in the paper. 

The best variational choice of $F_{ij}$ would be given by the 
free minimization of the FHNC/SOC energy and 
the solution of the corresponding Euler equations, 
$\delta E/ \delta F_{ij} = 0$. 
This approach is not practicable in finite nuclear systems,  
so we use an effective correlation obtained by the minimization of 
the energy evaluated at the lowest order of the cluster expansion, 
$E_{LO}$. The two--body Euler equations are then solved under 
the {\em healing} conditions $f^{1}(r\geq d_{1})=1$, 
$f^{p>1}(r\geq d_{p})=0$ and 
requiring that the first derivatives of these functions at $r=d_p$ vanish. 
The healing distances $d_{p}$ are taken as variational
parameters. In analogy with the nuclear matter case, 
in our calculations we adopt only two healing distances:
$d_c$ for the four central channels and
$d_t$ for the two tensor ones. 
Additional variational parameters are the quenching 
factors, $\alpha^p$, of the NN potentials in the Euler equations. 
More details are given in Ref. \cite{pan79} for nuclear matter 
and in Ref. \cite{fab98} for finite nuclei. 
In the CMC calculation of Ref. \cite{pie92} a nuclear 
matter Euler correlation was used and the nuclear matter Fermi 
momentum, $k_F$, was used as a variational parameter. 
We follow here the same strategy.

The single particle wave functions, $\phi_\alpha(i)$, used in this work 
have been obtained either by solving the single particle
Schr\"odinger equation with
Woods--Saxon potential, 
\begin{equation}
V_{WS}(r)= {V_0 \over {1+exp[(r-R_0)/a_0]}} ,
\label{WS}
\end{equation}
or with a harmonic oscillator well, $V_{HO}(r)$, with oscillator length 
$b_{HO}=\sqrt {\hbar /m\omega}$. The parameters $V_0$, $R_0$, and $a_0$ 
of $V_{WS}(r)$ and $b_{HO}$ of $V_{HO}(r)$ are also variationally 
determined.

It is possible to express
the expectation values of $n$--body operators in terms 
of $n$--body density matrices, and related quantities. 
In particular, the one- and two-body densities, $\rho_1({\bf r}_1)$  and 
$\rho_2^p({\bf r}_1,{\bf r}_2)$ , defined as:
\begin{equation}
\rho_1({\bf r})=
\langle \sum_{i} \delta({\bf r} - {\bf r}_i) 
 \rangle 
\label{rho1}
\end{equation}
and 
\begin{equation}
\rho_2^p({\bf r},{\bf r}')=
\langle \sum_{i\neq j} \delta({\bf r} - {\bf r}_i) \delta({\bf r}' 
- {\bf r}_j) 
O^p_{ij} \rangle ,
\label{rho2}
\end{equation}
are needed to compute the energy mean value (\ref{enfun}). 
In the above expressions we indicate the mean
value of an operator $Q$ as: 
$ \langle Q\rangle=\langle \Psi|Q|\Psi\rangle/
\langle\Psi|\Psi\rangle$.
 Cluster expansion and Fermi HyperNetted
Chain theory provide a viable way to evaluate the densities 
both in infinite and finite Fermi systems. 
We present here only some of the basic features of the FHNC/SOC
computational scheme. 
More complete discussions of the FHNC theory 
are found in Refs.\cite{ros82,co92} for scalar  
correlations and in Refs.\cite{pan79,fab98} for the operatorial case. 

In FHNC theory, with scalar correlations,
the densities are written in terms of the
correlation, $f^1(r)$, and of the {\em nodal} and 
{\em elementary} functions, $N^1_{xy}({\bf r}_1,{\bf r}_2)$ 
and $E^1_{xy}({\bf r}_1,{\bf r}_2)$, representing the sums of 
the corresponding nodal and elementary cluster diagrams. 
The $x$($y$) label characterizes the exchange pattern 
at the external points 1(2) and can indicate: a 
direct link, ($d$), if the particle is not exchanged and the point is
reached by a dynamical correlation, $h(r)=[f^1(r)]^2-1$;
an exchange link, ($e$), if the particle belongs to a closed exchange 
loop; a cyclic link, ($c$), if the particle belongs to an open  
exchange loop. 
The possible combinations are: $dd$, $de$, $ed$, $ee$ and $cc$. 
In infinite, homogeneous systems, as nuclear matter, the FHNC functions 
depend only on the inter-particle distance, $r_{12}$. The FHNC 
integral equations allow for computing the nodal functions, once 
the elementary ones are known.

With the introduction of operatorial correlations, the 
nodal functions gain a state dependence, 
$N^{p\geq 1}_{xy}({\bf r}_1,{\bf r}_2)$. The correlation operators 
do not commute anymore among themselves, $[F_{12},F_{13}]\neq 0$, 
and one must take into account the various possible orderings in 
(\ref{G}). 
For this reason a complete FHNC treatment for the full, state
dependent densities is, at present, not possible. 
The Single Operator Chain  approximation  was introduced in 
Ref.\cite{pan79} for nuclear matter, and extended to {\em ls} doubly 
closed shell nuclei in Ref. \cite{fab98}, for $f_6$ correlations. 
The SOC scheme 
consists in summing those $p>1$ chains, where each 
link may contain only one operatorial element and scalar dressings 
at all orders. Notice that operatorial dependence comes also on 
account of the exchanges of two or more  nucleons, since the space 
exchange operator is given by  
$P_{ij}=-\sum_{p=c,\sigma,\tau,\sigma\tau}O^p_{ij}/4$.
Here, and in the following, we may refer to
the operatorial channels as $c$ ($p=1$), $\sigma$ (spin), 
$t$ (tensor) and $b$ (spin--orbit). The isospin channels have an 
extra $\tau$--label.

The elementary functions, $E_{xy}({\bf r}_1,{\bf r}_2)$, 
represent an input to the FHNC equations, 
as they cannot be calculated in a closed form. 
The FHNC/0 approximation consists in neglecting all the 
elementary contributions. 
This seemingly crude approximation is actually based on 
the fact that the elementary diagrams are highly connected and have, 
at least, a  quadratic dependence on the density of the system.
These diagrams are not expected to produce relevant contributions in
the relatively low density nuclear systems, 
whereas they are important in denser systems, like atomic liquid Helium. 
A test of the validity of the 
FHNC/0 approximation, and, in general, of the importance of the
elementary diagrams,
is provided by the degree of accuracy in satisfying the densities 
normalizations. 
In Ref. \cite{fab98} particular attention 
has been  paid to the normalization of the one-body density,
\begin{equation}
A=\int d^3r_1 \rho_1({\bf r}_1)
\label{SR1}
\end{equation}
and to that of the central and isospin two-body densities,
\begin{equation}
1={{1}\over{A(A-1)}}\int d^3r_1\int d^3r_2 
\rho_2^c({\bf r}_1,{\bf r}_2), 
\label{SR2}
\end{equation}
\begin{equation}
-1={{1}\over{3A}}\int d^3r_1\int d^3r_2 
\rho_2^{\tau}({\bf r}_1,{\bf r}_2) . 
\label{SRtau}
\end{equation}

Deviations of the sum rules from their exact values are due to ({\em i}) 
the approximate evaluation of the elementary diagrams and 
({\em ii}) the SOC approximation. 
The first item has been already investigated 
in Ref.\cite{co92}, where it has been found that the most relevant 
corrections to the FHNC/0 sum rules come from the $E_{ee}^{exch}$ 
diagrams, {\em i.e.} $ee$-elementary diagrams whose external points 
belong to the same exchange loop. 
These diagrams mainly contribute to the isospin 
saturation sum rule (\ref{SRtau}) and to the potential energy, if 
the interaction has large exchange terms. 
This is understood if we 
notice that a four-point elementary diagram, linear in the central link 
$h(r)$, belongs to $E_{ee}^{exch}$, as well as 
diagrams linear in the operatorial link, $f^1(r)f^{p>1}(r)$. 

In Ref. \cite{fab98} it has been shown that the 
one--body density sum rule is violated in
FHNC/SOC  by  less than 1$\%$ 
and the two-body density normalizations (\ref{SR2},\ref{SRtau}) by 
$\sim$9$\%$ in the worst case ($^{40}$Ca with tensor correlations). 
This is the same degree of accuracy found  in
Ref.\cite{wir88} for nuclear matter calculations.

In Ref. \cite{fab98} the comparison of the
$^{16}$O FHNC energies calculated for a purely Jastrow correlation
with the exact VMC estimates shows an excellent agreement,  
with a difference of $\sim$1$\%$ in the kinetic energy 
( 24.61 MeV in FHNC vs. 24.33$\pm$0.21 in VMC) 
and of $\sim$2$\%$ in the potential energy  
( -22.07 MeV in FHNC vs. -21.56$\pm$0.25 in VMC).
The $^{16}$O FHNC energies calculated with a  $f_6$ correlation and
using the $v_6$ truncation of the Urbana $v_{14}$ 
potential \cite{lag81}, have been compared with the results of a
fifth order CMC calculation, which appeared to have reached a 
satisfying convergence. In this case
we found a difference of less than 5$\%$ for the kinetic energy 
( 31.16 MeV in FHNC/SOC vs. 29.45$\pm$0.33 in CMC) and of $\sim$7$\%$ 
for the potential one 
( -35.47 MeV in FHNC/SOC vs. -33.03$\pm$0.31 in CMC). 
The FHNC/SOC calculation gives a binding energy per nucleon of
-4.33 MeV to be compared with the -4.59$\pm$0.10 MeV value 
obtained by CMC.  
This difference is compatible with the nuclear matter estimates.

\section{Spin--Orbit and three--body forces}

As already mentioned in the introduction, the novelty of this 
work with respect to Ref.\cite{fab98} is the inclusion of the 
spin--orbit and three--body terms of the potential. 
In this section we briefly show how to extend
the FHNC/SOC formalism to consider these parts of the interaction. 

\subsection{Spin--Orbit potential}

The treatment of the spin-orbit interaction within the FHNC/SOC
formalism has been discussed in details in
Ref.\cite{lag80} for the nuclear matter case. 
In that paper, the evaluation of the mean value of the
spin-orbit terms of the interaction ,
$\langle v^{7\equiv b,8 \equiv b\tau} \rangle$, 
was done using a $f_8$ correlation factor, so including 
all the 8 operatorial components in Eq.(\ref{corr8}).

Here we extend the nuclear matter
formalism to the finite nuclei case for a $f_6$ correlation. 
A correlation containing spin--orbit components would,
probably, be more efficient from the variational point of view.  
However, these correlation terms introduce extra uncertainties 
in the cluster expansion, and we have chosen to work with a 
simpler correlation in favor of a safer convergence. 
We shall discuss further this point and
give some estimates of the $f^{p>6}$ corrections.

In the FHNC/SOC scheme, the 
ground state matrix element, $W$, of a generic two--body 
operator, $\hat W$, is split into four parts
\begin{equation}
W=W_0+W_s+W_c+W_{cs}\,\,.
\label{wmatr}
\end{equation}
where $W_0$ indicates the sum of the diagrams with central chains 
between the fully correlated interacting points, 
connected by $\hat W$, 
$W_s$ the sum of the diagrams having
operatorial vertex corrections (or Single Operator Rings) touching the
interacting points and central chains, 
$W_c$ the diagrams with one operatoral 
chain, SOC, between the interacting points, and, finally,
$W_{cs}$ the diagrams containing both operatorial 
vertex corrections and chains. 
A more complete discussion on this topic is found in
Refs.\cite{pan79,fab98}.  

The nuclear matter calculations of
Ref.\cite{lag80} show that two--body clusters provide
the leading contribution to $\langle v^{b,b\tau} \rangle$. 
Three body separable diagrams, contributing to $W_s$, gave the 
main many--body contributions and were of the order of 10$\%$ of the 
two body ones. Chain contributions were even 
smaller. These results  have been confirmed in Ref.\cite{akm97}, 
where the authors found 
$\langle v^{b+b\tau} \rangle_{two-body}/A=$-2.29 MeV  and a 
many-body contribution ($W_s+W_c$) of -0.28 MeV at the nuclear matter 
saturation density, with the A18 potential.  

Relying on these facts, for the spin--orbit terms of the interaction,
we have calculated only the $W_0$ contribution
of Eq.(\ref{wmatr}) with a $f_6$ correlation function. 
It turns out that, in this case,
only the tensor correlations contribute, and the full expression 
reads as:
\begin{eqnarray}
&\langle v^{bj_\tau}\rangle_{W_0}&=-9
\int d^3r_1 \int d^3r_2 f^{ti_\tau}
(r_{12})v^{bj_\tau}(r_{12})f^{tk_\tau}(r_{12}) 
h^c({\bf r}_1,{\bf r}_2)
 \\ \nonumber & & \left\{
 \rho^c({\bf r}_1)\rho^c({\bf r}_2)  K^{i_\tau j_\tau k_\tau}A^{k_\tau}    
 - 8 (N_{cc}^c ({\bf r}_1,{\bf r}_2) -\rho_0({\bf r}_1,{\bf r}_2))^2  
 K^{i_\tau j_\tau l_\tau }K^{l_\tau k_\tau m_\tau} 
A^{m_\tau}\Delta^{m_\tau} C_d({\bf r}_1) C_d({\bf r}_2) \right \} . 
\label{w0matr}
\end{eqnarray}
In the above equation 
a sum over repeated indexes is understood, 
($i_\tau ,j_\tau ,k_\tau,l_\tau,m_\tau=c,\tau$), 
the coefficients $K^{ijk}$, $A^i$   
and $\Delta^i$ are given in Ref.\cite{pan79} and the other FHNC functions are 
defined in \cite{fab98}. 

For the $^{16}$O nucleus, we could check 
the reliability of this approximation against 
some CMC results \cite{pie99}. 
Using a nuclear matter correlation and the A14 potential, we find 
$\langle v^{b+b\tau} \rangle_{W_0}/A=$ 0.56 MeV 
whereas CMC, with the same correlation, gives an
extrapolated value of 0.62 MeV. We did not find 
the same kind of agreement for a $f_8$ correlation, where 
the $W_0$ result is -1.23 MeV and CMC gives 0.03 MeV. This supports 
our choice in favor of the use of a simpler $f_6$ correlation.

\subsection{Three--body potential}

The evaluation of the mean value of the three-body interaction,
$\langle v_{ijk}\rangle=
\langle v^{2\pi}_{ijk}\rangle+\langle v^R_{ijk}\rangle$ 
closely follows 
the method developed in Ref.\cite{car83} for nuclear matter.

Diagrams 2.1--3 of the reference were shown to provide  the 
relevant contributions to $\langle v^{2\pi}_{ijk}\rangle$, and 
diagrams 3.1 and 3.2 to $\langle v^R_{ijk} \rangle$. These diagrams, 
which we show in Figure \ref{fig:diag}, are those considered in our
calculations.

As an example, we show here how the nuclear matter expression 
of diagram 2.1 is extended to our case. 
The explicit expressions for the remaining 
terms are given in the Appendix.

In diagram 2.1 the pairs of nucleon connected by the 
operators $X_{ij}$ (pairs 12 and 13 in the figure) 
are dressed at all orders by Jastrow correlations, whereas the 
remaining pair (23) bears the full operatorial correlations. 
Only the anticommutator part of $v^{2\pi}_{ijk}$ contributes 
to $\langle v^{2\pi}_{ijk}\rangle_{2.1}$. Spin-ispospin trace 
and spatial integration over nucleon 1 generate an effective 
two--body potential, acting on the 23 pair, having $\sigma\tau$ and 
$t\tau$ components only and depending on the exchange patterns, 
($xy$), of particles 2 and 3, 
\begin{equation}
v^{eff}_{xy}({\bf r}_2,{\bf r}_3) = \sum_{k=\sigma\tau,t\tau}
v^{eff,k}_{xy}({\bf r}_2,{\bf r}_3)O^k_{23},
\label{Veff_1}
\end{equation}
with
\begin{equation}
v^{eff,k}_{xy}({\bf r}_2,{\bf r}_3)= 4A_{2\pi}
\int d^3r_1 
\left[ \sum_{x'y'}
 g^c_{xx'}({\bf r}_2,{\bf r}_1)
V^c_{x'y'}({\bf r}_1)
 g^c_{y'y}({\bf r}_1,{\bf r}_3) \right] 
\xi^{i_\sigma j_\sigma k_\sigma }_{213} 
X^{i_\sigma}(r_{21})X^{j_\sigma}(r_{13}) ,
\label{Veff_2}
\end{equation}
where $k=k_\sigma \tau$, 
($i_\sigma$, $j_\sigma$, $k_\sigma$= $c$, $\sigma$, $t$),  
$g^c_{xy}$ and $V^c_{xy}$ are central 
partial distribution functions and vertex corrections defined 
in \cite{fab98}, $\xi^{ijk}_{213}$ are angular couplings given in 
Ref.\cite{pan79} and $X^{c}(r)=0$, $X^{\sigma}(r)=Y(r)$ and 
$X^{t}(r)=T(r)$.  
The allowed  combinations for ($x'y'$) are: $dd$, $de$, $ed$, and 
$cc$. In the last case, also ($xy$) should be ($cc$). 

The full expression for the mean value of the 2.1 diagram,
$\langle v^{2\pi}_{ijk}\rangle_{2.1}$, is then given by:
\begin{eqnarray}
&\langle v^{2\pi}_{ijk}\rangle_{2.1} &={{1}\over{2}}
\int d^3r_2 \int d^3r_3 f^i(r_{23}) f^j(r_{23}) 
h^c({\bf r}_2,{\bf r}_3) 
K^{ikj}A^j 
 \\ \nonumber & & 
\left \{ v^{eff,k}_{dd}({\bf r}_2,{\bf r}_3) 
\left[ \rho^c_1({\bf r}_2)\rho^c_1({\bf r}_3) + 
\rho^c_1({\bf r}_2) N^c_{de}({\bf r}_2,{\bf r}_3) C_d({\bf r}_3) + 
C_d({\bf r}_2) N^c_{ed}({\bf r}_2,{\bf r}_3) \rho^c_1({\bf r}_3) 
 \right .\right .
 \\ \nonumber & & 
\left . + 
C_d({\bf r}_2) 
\left( N^c_{de}({\bf r}_2,{\bf r}_3) N^c_{ed}({\bf r}_2,{\bf r}_3) + 
 N^c_{ee}({\bf r}_2,{\bf r}_3) \right) C_d({\bf r}_3) \right] 
 \\ \nonumber & & 
 + v^{eff,k}_{de}({\bf r}_2,{\bf r}_3) 
\left[ \rho^c_1({\bf r}_2) + 
C_d({\bf r}_2) N^c_{ed}({\bf r}_2,{\bf r}_3) \right] C_d({\bf r}_3) 
 \\ \nonumber & & 
 + v^{eff,k}_{ed}({\bf r}_2,{\bf r}_3) C_d({\bf r}_2)
\left[ \rho^c_1({\bf r}_3) + 
 N^c_{de}({\bf r}_2,{\bf r}_3)C_d({\bf r}_3) \right]    
 \\ \nonumber & & \left . 
 + v^{eff,k}_{ee}({\bf r}_2,{\bf r}_3) C_d({\bf r}_2)C_d({\bf r}_3)
\right\}    
 \\ \nonumber & & 
- 2
\int d^3r_2 \int d^3r_3 f^i(r_{23}) f^j(r_{23}) 
h^c({\bf r}_2,{\bf r}_3) C_d({\bf r}_2)C_d({\bf r}_3)
K^{ikl} K^{ljm} A^m \Delta^m 
 \\ \nonumber & & 
\left \{ v^{eff,k}_{dd}({\bf r}_2,{\bf r}_3) 
\left[ N^c_{cc}({\bf r}_2,{\bf r}_3) -\rho_0({\bf r}_2,{\bf r}_3) \right]^2 
+ 2 v^{eff,k}_{cc}({\bf r}_2,{\bf r}_3) 
\left[ N^c_{cc}({\bf r}_2,{\bf r}_3) -\rho_0({\bf r}_2,{\bf r}_3) 
\right] \right\} .
\label{TNI_21}
\end{eqnarray}

Both commutator and anticommutator contributions are present in 
$\langle v^{2\pi}_{ijk}\rangle_{2.2}$, while only the commutator 
part contributes to $\langle v^{2\pi}_{ijk}\rangle_{2.3}$. 

In Table \ref{tab:3body} we compare 
the TNI mean values for $^{16}$O calculated by FHNC/SOC 
with the  MonteCarlo estimates \cite{pie99} obtained with the 
same wave function. 
We have used the Urbana V three-nucleon interaction \cite{car83} and the 
$v_6$ truncation of the Urbana $v_{14}$ NN interaction. 
The correlations have been obtained by the two--body Euler equation and
we used harmonic oscillator single particle states with $b=1.54$ fm.
The $^{40}$Ca results are also shown.

For the Jastrow correlation the MonteCarlo results have been obtained
with a VMC calculation, while for the $f_6$ correlation the
calculation is of the CMC type.
In $^{16}$O the FHNC/SOC results are in satisfactory agreement with the
MonteCarlo ones for both the classes of correlations.
The $^{40}$Ca results may be compared with those in 
nuclear matter, where, at saturation density, the Urbana V model 
gives  $\langle v^{2\pi}_{ijk}\rangle$=-2.32 MeV and 
$\langle v^{R}_{ijk}\rangle$=3.35 MeV\cite{car83}.

The change of sign of  $\langle v^{2\pi}_{ijk}\rangle$ between
the Jastrow and the operatorial correlation is due to the fact that 
the attractive contribution comes most from the tensor component of the 
effective potential, $v^{eff,t\tau}_{xy}({\bf r}_2,{\bf r}_3)$, which 
does not contribute in absence of tensor correlations.

\section{Results}

The results presented in this section have been obtained
with $v_8$ type NN interactions, either by truncating the A14
potential \cite{wir84}  or by using the A8' potential of
Ref. \cite{pud97}, together with $f_6$ correlations and 
Urbana three--nucleon potentials.

We have estimated  the $f^{p>6}$ corrections, as well 
as the contributions from the $p>8$ components of the potential, 
by  Local Density Approximation (LDA).
In practice, if we define as $\Delta E_{nm}(\rho)$ 
the sum of these corrections in nuclear matter at density $\rho$, 
we evaluate their contribution in the finite nucleus as: 
\begin{equation}
{\Delta E}= {1\over A} \int d^3r \rho_1({\bf r}) \Delta E_{nm}
[\rho_1({\bf r})].
\label{LDA}
 \end{equation}

We have already studied the accuracy of FHNC/SOC 
computational scheme against the results of CMC in Ref. \cite{fab98}. 
In the present article we compare again 
the $^{16}$O results with those of CMC in order to test the
accuracy in the calculation of the interaction terms we 
have added. For this reason
we have computed the ground state energy for the A14+UVII model 
using a $f_6$ correlation derived from the nuclear 
matter two--body Euler equations. The parameters of the correlation 
are: the Fermi momentum $k_F$, and the healing distances $d_c$, used for
all the central channels, and $d_t$ for the tensor channels.
Additional variational parameters are the quenching factors 
$\beta_p$ of the potential (see eq. 3.2 of \cite{pan79}), and
as in Ref. \cite{fab98} we have taken $\beta_1=1$ 
and $\beta_{p>1}=\alpha_S$.
We have used the same set of single particle states, produced by 
a Woods--Saxon plus wine--bottle mean field potential, and correlation 
parameters of Ref. \cite{pie92}. However, small differences in the 
correlation come on account of the fact that our nuclear matter 
Euler equation does not contain the $v^{p>6}$ components, contrary 
to Ref. \cite{pie92}.

The results of these calculations are shown in Table \ref{tab:cmc}.
The Table gives also the corresponding CMC results, 
as extracted from the reference, by subtracting, 
when reported, the three--body and spin--orbit  correlations 
contributions.
The $\Delta E_{CMC}$  line gives the CMC 
corrections coming from $(f,v)^{p>6}$ and should be compared with 
the $\Delta E+\langle v_{7-8,ij}\rangle$ sum 
(-0.5 vs. -0.80 MeV/A) in the FHNC/SOC column.
The ground state energy, $E_{gs}$, is given by 
$E_{gs}=E_v-T_{cm}$, where $T_{cm}$ is the 
center of mass kinetic energy. The Coulomb potential energy, 
$\langle v_{Coul}\rangle$, has been included in $E_{gs}$ 
The root mean square radii, $rms$, are also reported. 

The energy found in CMC is -7.7 MeV/A, and it contains a -0.85 MeV/A
contribution from explicit three--body correlations. 
Therefore, the FHNC/SOC result 
(-5.97 MeV/A) should be compared with an estimated CMC value
of -6.85 MeV/A. 
If the CMC expansion has reached a satisfactory convergence, 
the discrepancies between the two calculations are due to the 
truncation in the FHNC/SOC scheme and to the aforementioned differences 
in the correlation. As it was found in Ref.\cite{fab98}, 
the FHNC ground state energy differs from the CMC estimate by 
$\sim$1 MeV/A, compatible with the estimated accuracy of the method 
in nuclear matter. In this calculation the
$\Delta E$ corrections are small, somehow justifying a posteriori the 
use of LDA.  
It is remarkable that
in the coupled cluster approach of Ref.\cite{hei99} 
the authors find $E_{gs}$(CC)=-6.1 MeV/A with 
the two--body A14 model, 
close to our estimated $E_{gs}$(FHNC/SOC)=-6.04 MeV/A (obtained
without $<v_{ijk}>$ and $<v_{Coul}>$).

The nuclear matter energies per nucleon calculated with
several interactions are given in Tab. \ref{tab:nm} as a 
function of the density.
The A18+UIX (A18) and  A8'+UIX (A8') interactions
provide energy minima at the empirical value of the saturation density.
The results of the A18 column differ from those of Ref.\cite{akm98} 
 because  we have subtracted a perturbative correction, related to 
additional state dependence of the correlation.
The second column of the Table shows the results for the 
A8'+UIX interaction with a $f_8$ correlation. The contribution of the 
$f^{p > 6}$ terms is given in the third column.
 For completeness, we also give the results obtained with the full
A14 potential plus the UVII three--body force and the corresponding 
$\Delta$E/A. The A14+UVII (A14) minimum of the equation of 
state is located at a density slightly higher than the empirical one.

The ground state energies of
$^{16}$O and $^{40}$Ca calculated with 
the A8'+UIX and the truncated A14+UVII
interactions are shown in Table IV. 
We have adopted the nuclear matter $f_6$ correlation, where 
the nuclear matter density is used as a variational parameter 
(it means that, for a given $\rho_{nm}$, we use the correlation 
function parameters minimizing the nuclear matter energy at that 
density). In addition, 
the energy has been minimized over the single particle potential 
(HO or WS) parameters. 
%
%
 The Table also contains the kinetic energies computed with 
 only the mean field wave functions. They are about half of the 
 total kinetic energies  obtained in the full calculations 
 and this difference has to be ascribed to the correlations. 
 This type of behavior is also found in nuclear matter. 
 For example, at saturation density and with the A18+UIX hamiltonian, 
 the nuclear matter calculations of Ref.\cite{akm98} provide 
 a total kinetic energy of 42.27 MeV/A to be compared with the
 corresponding  Fermi gas value of 22.11 MeV/A.
%
%
As expected, the minimization with the WS potentials produces lower
minima than those obtained with HO potentials.
The comparison between the WS calculations done with the two different
interactions shows small differences in the results, of the order of
7\% in $E_{gs}$. From the relative point of view the largest
differences are those related to $\Delta E$, which for A14 are 50\%
smaller than those of A8'. 
These rather similar values of the energies have been obtained with
two  quite different sets of single particle wave functions.
This can be deduced by looking at the values of the WS parameters in
Tab. \ref{tab:hows}, 
and even better from Figure \ref{fig:dens6} where the FHNC/SOC
charge distributions are compared with the IPM ones and with the
empirical densities taken from the compilation of Ref. \cite{dej87}.
The charge densities have been obtained by folding the proton
densities with the electromagnetic form factors of Ref. \cite{hoe76}.

%
%


 We do not obtain a satisfactory agreement between the computed  
 densities and the empirical ones. However, we remark that, for 
 a given type of single particle wavefunctions, either HO or WS, 
 we find shallow energy minima with respect to variations of the
 mean field parameters around the minimum itself. This may indicate 
 that charge distributions and rms radii are sensitive to details 
 of the many body wave function which have small effects on the 
 energy calculation.
%
%
To better study this aspect, we have done calculations by using a set of  
single particle wave functions fixed to reproduce at best the
empirical charge densities. In this case, the only variational parameters
are those related to the correlation functions. Actually, we have 
varied only $d_t$ and kept fixed its ratio with  $d_c$.
The results of these calculations with the A8'+UIX  interaction 
are shown in Tab. \ref{tab:wsfit} and in Figure \ref{fig:dens2}. 
%
%
 We observe that a large  change in the single particle wave
 functions produces small variations in the energy values, 
 1.3\% in $^{16}$O and 4.8\% in $^{40}$Ca, within the accuracy of
 the FHNC/SOC scheme. The analysis of these results 
 shows that the values of the kinetic and potential energies have
 considerably changed, while their sum remains almost constant.
%
%
The agreement with the empirical densities has clearly improved, 
as one can also see in Figure \ref{fig:cross} where the 
elastic electron scattering cross sections calculated in 
Distorted Wave Born Approximation \cite{ann95}
with the FHNC/SOC charge densities are
compared with the experimental data \cite{sic70}.
The  best agreement with the data is produced by the 
densities of Figure \ref{fig:dens2}. 
However, also in this case the high momentum and large energy data 
are not well described in both nuclei.

 From the comparison between the dashed and the thin full lines in 
Figures \ref{fig:dens6} and \ref{fig:dens2} 
we can inspect the effects of the correlation on the charge densities.
In $^{16}$O the correlations enhance the densities with respect 
to the IPM ones without substantially changing the shape of the 
oscillation. The effect of the correlations on the $^{40}$Ca 
densities is negligible.
These results confirm the findings of Ref. \cite{ari97}
where the charge densities have been calculated up to the first order
in the correlation lines. On the other hand, in Ref. \cite{pie92} the
correlations produce much larger deviations from the IPM densities. 
%
%
 The reason of the different effect of the correlations on the densities
 of $^{40}$Ca and $^{16}$O is presently object of investigation.
%
%

Short range correlations strongly affect the
two--body densities (\ref{rho2}). 
Their effect is evident in the 
two--nucleon distribution function, 
$\rho_2(r_{12})$, defined as:
\begin{equation}
\rho_2( r_{12})=\frac {1}{A}\int d^3 R_{12}
\rho_2^c({\bf r}_1,{\bf r}_2)
\label{2NDF}
\end{equation}
where ${\bf R}_{12}=\frac {1}{2}\left( {\bf r}_1 + {\bf r}_2 \right)$ 
is the center of mass coordinate. 
In an analogous way,  we define the 
proton--proton distribution function, $\rho_{pp}(r_{12})$, as:
\begin{equation}
\rho_{pp}( r_{12})=\frac {1}{Z}\int d^3 R_{12}
\rho_{pp}({\bf r}_1,{\bf r}_2), 
\label{ppDF}
\end{equation}
where the $pp$--two--body density is: 
\begin{eqnarray}
&\rho_{pp}({\bf r}_1,{\bf r}_2)&=
\langle \sum_{i\neq j} 
\delta({\bf r}_1 - {\bf r}_i) \delta({\bf r}_2 - {\bf r}_j) 
\left({{1+\tau_{i,z}}\over 2}\right)
\left({{1+\tau_{j,z}}\over 2}\right) \rangle 
 \\ \nonumber & & 
={1\over 4} \rho_2^c({\bf r}_1,{\bf r}_2)
+{1\over {12}} \rho_2^\tau({\bf r}_1,{\bf r}_2) .
\label{rhopp}
\end{eqnarray}

Figure \ref{fig:tbd} shows 
$\rho_{2}(r_{12})$ and $\rho_{pp}( r_{12})$ with the wave functions 
of Tab. \ref{tab:wsfit}, compared with the IPM densities. 
The reduction of the 
correlated distribution functions at small $r_{12}$--values is due 
to the repulsive core of the interaction. 
We have calculated the FHNC/SOC $^{16}$O distribution 
functions also for the A14+UVII model and we found that they are
rather similar to the  A8'+UIX ones 
($\rho_2^{max}(A14+UVII)\sim$0.081, 
$\rho_2^{max}(A8'+UIX)\sim$0.089, $\rho_{pp}^{max}(A14+UVII)\sim$0.032 
and $\rho_{pp}^{max}(A8'+UIX)\sim$0.032), and in good agreement 
with those computed in Ref.\cite{pie92}.

The interest in calculating the two--body densities
is also related to the possibility of using their Fourier 
transforms to analyze several integrated nuclear responses.
The responses of the nucleus to external probes, 
either of electromagnetic or hadronic type, 
can be related to the dynamical structure 
functions, $S_X(q,\omega)$, given by
\begin{equation}
S_X(q,\omega)=
 \sum_I \vert \langle \Psi_I \vert O_X \vert \Psi_0 \rangle \vert ^2 
\delta ( \omega_I - \omega_0 - \omega ) , 
\label{Sx_qw}
\end{equation}
where $O_X$ is the operator producing the fluctuations around the  
ground state $\Psi_{0}$. In the above equation the sum runs over the
intermediate $\Psi_{I}$--states with energy $\omega_{I}$. 
The non energy weighted sums of $S_X(q,\omega)$ give 
the static structure functions (or, simply, structure functions, SF), 
$S_X(q)$, as:
\begin{equation}
S_X(q)= \int S_X(q,\omega) d\omega = 
 \langle \Psi_0 \vert O_X^\dagger O_X \vert \Psi_0 \rangle . 
\label{Sx_q}
\end{equation}

In the case of density fluctuations, the operator is:
\begin{equation}
\rho_{\bf q}= \sum_{i=1,A} \exp ( \imath {\bf q}\cdot {\bf r}_i )
\label{rho_q}
\end{equation}
and the lower limit $\omega$--integration in (\ref{Sx_q}) is taken 
in an appropriate way to eliminate the contribution of the 
elastic scattering. 
The density SF, $S(q)$, is then 
\begin{equation}
S(q)= 1 + {1\over A} \int d^3r_1 d^3r_2
 \exp ( \imath {\bf q}\cdot {\bf r}_{12} )
\left\{\rho_2^c({\bf r}_1,{\bf r}_2)-
\rho_1({\bf r}_1) \rho_1({\bf r}_2)\right\} .
\label{S_q}
\end{equation}
 From the normalizations of the one-- and two--body densities, one 
obtains $S(q=0)=$0.  

The response to charge fluctuations is driven by the operator
\begin{equation}
\rho_{c,\bf q}= \sum_{i=1,A} \exp ( \imath {\bf q}\cdot {\bf r}_i )
\left({{1+\tau_{i,z}}\over 2}\right) ,
\label{rho_cq}
\end{equation}
which is responsible also of the electromagnetic longitudinal 
response, if the small ($\sim 2\%$) contributions of the neutron 
magnetic moment and of the meson--exchange currents are disregarded. 
The longitudinal response is measured in inelastic
electron--nucleus experiments  
and its energy integral gives the longitudinal SF, $S_L(q)$, or 
Coulomb sum. In the nuclei we are studying, the explicit expression 
of $S_L(q)$ reads: 
\begin{equation}
S_L(q)= 1 + {1\over {4Z}} \int d^3r_1 d^3r_2
 \exp ( \imath {\bf q}\cdot {\bf r}_{12} )
\left\{\rho_2^c({\bf r}_1,{\bf r}_2)+
{1\over 3}\rho_2^\tau({\bf r}_1,{\bf r}_2)-
\rho_1({\bf r}_1) \rho_1({\bf r}_2)\right\} .
\label{Sc_q}
\end{equation}

In Figure \ref{fig:sfsl} we present the  density (upper panel) and charge
(lower panel) SF for $^{16}$O, calculated with the wave functions 
of Tab. \ref{tab:wsfit}.
The FHNC/SOC results obtained from our calculations are shown by the
white triangles. 
The $q=0$ value is very sensitive even to small violations of the 
normalizations of the densities.
For instance, the two--body density normalization of Eq.(\ref{SR2}),  
with the Tab. \ref{tab:wsfit} parameters, 
is violated in $^{16}$O by only 5.3$\%$ 
and that of the one--body density, Eq.(\ref{SR1}), by 2.0$\%$. 
These acceptable normalization errors produce the large values $S(q=0)$=1.61 
and $S_L(q=0)$=0.86. 

The SF obtained after properly renormalizing the densities are given in 
Figure \ref{fig:sfsl} as full curves. 
The other curves show the
IPM SF (dashed lines) and those obtained 
with only Jastrow correlations. The renormalization is 
effective only at small $q$--values and for  $q>$1 fm$^{-1}$ 
the SF remain unchanged. Analogous calculations done in $^{40}$Ca
show a similar behavior. Hence, the conclusions about the large--$q$
importance of the correlations in the SF and densities are not 
affected by the normalization problems. 
In agreement with the findings of Ref. \cite{pie92}, our results show
that the correlations, both of the Jastrow and operatorial types, 
 lower the SF at large $q$.

In Figure \ref{fig:csr} we compare the Coulomb sum rules calculated
for $^{16}$O and $^{40}$Ca with the  A8'+UIX
interaction with the experimental estimates done by analyzing
the set of
world data on inclusive quasi--elastic electron scattering \cite{jou96} 
experiments in $^{12}$C, $^{40}$Ca, and $^{56}$Fe. 
The Figure also shows the nuclear matter Coulomb sum for the 
A14+UVII model from Ref. \cite{sch87}. 
The finite nuclei results are in complete 
agreement with the  latest analysis of the experimental data, where a 
detailed study of the electron scattering world experiments and a proper 
inclusion of the large energy tail in the dynamical response were 
carried out.
The nuclear matter results fail in reproducing the data at the
lowest $q$ values where finite size effects can be still important.
 
Besides the density and charge SF also
the isovector spin longitudinal and transverse (ISL and IST) SF are
of experimental interest since they can be extracted from polarized 
proton and neutron scattering cross section data.
Experiments of this type on  $^{12}$C and $^{40}$Ca nuclei \cite{mcc92}  
did not confirm the prediction 
of Random Phase Approximation \cite{alb82} and 
Distorted--Wave Impulse Approximation \cite{ich89} 
calculations of a large enhancement, with respect to unity, 
of the ratio of the ISL to the IST response at small energies.

The ratio has been calculated for nuclear matter within  
the CBF theory, using a $f_6$ correlation together with 
the Urbana $v_{14}$ +TNI potential \cite{fab94}. The computed 
average enhancement was $\sim 20\%$,
compatible with the data in heavy nuclei at energies below 100 MeV.
However, 
the nuclear matter calculation did not take into account the strong 
distortion of the emitted nucleon wave function.
Variational and cluster Monte 
Carlo wave functions were used in Ref.\cite{pan94} to evaluate 
the integrated spin responses in light nuclei and $^{16}$O. A 
maximum 25$\%$ enhancement of the ratio was found in $^{16}$O.     
 
The fluctuation operators in the isovector spin responses are
\begin{equation}
\rho_{\sigma L,\bf q}= \sum_{i=1,A} \exp ( \imath {\bf q}\cdot {\bf r}_i )
({\bf \sigma}_i \cdot {\bf q})
({\bf \tau}_i \cdot {\bf T})
\label{rho_sLq}
\end{equation}
in the longitudinal case, and 
\begin{equation}
\rho_{\sigma T,\bf q}= \sum_{i=1,A} \exp ( \imath {\bf q}\cdot {\bf r}_i )
({\bf \sigma}_i \times {\bf q})
({\bf \tau}_i \cdot {\bf T})
\label{rho_sTq}
\end{equation}
in the transverse one. 
In the above equations we have indicated with
${\bf T}$ a unit vector in the isospin 
space. If the nucleus has zero isospin, then the response does not 
depend on the direction of ${\bf T}$, except for the small Coulomb 
effects. 
Following the treatment of Ref.\cite{pan94}, we obtain for
the ISL structure 
function, $S_{\sigma L}(q)$, the expression: 
\begin{equation}
S_{\sigma L}(q) = 
{S^u_{\sigma L}(q)\over {Aq^2}} = 
1 + {1\over {9A}} \int d^3r_1 d^3r_2
\left\{\rho_2^{\sigma \tau}({\bf r}_1,{\bf r}_2) j_0(qr_{12})-
\rho_2^{t \tau}({\bf r}_1,{\bf r}_2)j_2(qr_{12}) \right\} ,
\label{SsL_q}
\end{equation}
and for the IST one, $S_{\sigma T}(q)$:   
\begin{equation}
S_{\sigma T}(q) = 
{S^u_{\sigma T}(q)\over {2Aq^2}} = 
1 + {1\over {9A} }\int d^3r_1 d^3r_2
\left\{\rho_2^{\sigma \tau}({\bf r}_1,{\bf r}_2) j_0(qr_{12})+
{1\over 2}\rho_2^{t \tau}({\bf r}_1,{\bf r}_2)j_2(qr_{12}) \right\} .
\label{SsT_q}
\end{equation}

The ISL and IST structure functions for $^{16}$O and $^{40}$Ca,
calculated with the wave functions of Tab. \ref{tab:wsfit},
are shown in Figure \ref{fig:spinsf} as a function of the momentum
transfer. 
For the $^{16}$O nucleus we show also the Jastrow results, 
identical for both ISL and IST. 
Central correlations do not differentiate between the longitudinal 
and transverse responses because of the lack of tensor correlations. 
Our $^{16}$O results are very close to those of  
Ref.\cite{pan94} obtained with the A14+UVII interaction. 
The lowest panel of Figure \ref{fig:spinsf} shows the  
$S_{\sigma L}(q)/S_{\sigma T}(q)$ ratio. We observe that 
the maximum enhancement is $\sim$25$\%$ in $^{16}$O and 
$\sim$38$\%$ in $^{40}$Ca,  close to the experimental value.

\section{Conclusions}

This paper is the natural extension of the work of Ref. \cite{fab98}.
We have added to the FHNC/SOC computational scheme for doubly closed
shell nuclei in $ls$ coupling, the contribution of the
spin--orbit and three--body interactions.
Our calculations of $^{16}$O and $^{40}$Ca nuclei have been done
considering a $f_6$ type of correlation and $v_8$ 
potentials: the A8'+UIX \cite{pud97} and a truncated version of the
A14+UVII \cite{wir84} potentials. The contribution from the 
remaining momentum dependent terms of the correlation and of the potential 
are estimated by means of a local density interpolation of their
nuclear matter values. 

The main results of the paper are given in Tabs. \ref{tab:hows}
and \ref{tab:wsfit}
where the $^{16}$O and $^{40}$Ca ground state energies per nucleon 
are presented. 
Their values are $\sim$ 2--3 MeV/A above the experimental ones, 
consistently with the CBF results in nuclear matter.
Additional lowering of the energies may be obtained by 
({\em i}) three--body correlations,  
({\em ii}) perturbative corrections to the two--body correlations. 
 It has been already mentioned that three--body correlations have been 
found to provide an extra $0.8$ MeV/A binding in $^{16}$O for the 
A14+UVII model. Perturbative corrections have 
been taken into account in nuclear matter either by the method 
discussed in Ref.\cite{akm98} or by the inclusion of the second order 
two--particle two--hole contribution in correlated basis perturbation 
theory \cite{fan83,fab93}. Both approaches lower the energy by 
$\Delta E_2\sim$2. MeV/A.  
The nuclear matter case gives a strong indication 
that the inclusion of these corrections 
should be pursued and that their quantitative consistency 
in finite nuclear systems  needs to be numerically checked.

A complete minimization over all the parameters of the wave functions,
both in the  correlation and in the mean field, has led to a marked 
disagreement between the CBF and the empirical charge densities in the low 
distance region. However, calculations with different sets of single 
particle wave functions, reproducing the empirical densities in IPM,  
provided energies differing from the best minima only by a few percent. 
The CBF scheme does not appear to be very sensitive 
to the details of the mean field basis in a parameters region around 
the variational minimum. 
Nevertheless, these details become
relevant for a correct description of the one-- and two--body densities. 
The introduction of additional constraints during the
minimization process may be necessary in order to avoid these ambiguities.

Our results show that the short--range correlations produce small
effects on the charge density distributions, especially in $^{40}$Ca
where the FHNC scheme is supposed to perform better. 
These findings are in
agreement with those of Ref. \cite{ari97} where the same kind of
nuclear matter correlations have been used. 
The sensitivity of the charge
distributions to the state dependent short--range correlations
requires further investigations to be fully clarified.
In effect, the VMC one body densities of \cite{pie92} have larger
dependence on these components.
 
In addition to the ground state energies, we have studied the static 
structure functions. 
The FHNC results for the Coulomb sum rule fully agree 
with the empirical values. 
The ratio between the ISL and IST SF shows an enhancement between 
25\% in $^{16}$O and 38\% in $^{40}$Ca, in agreement with those of
 Refs. \cite{fab94,pan94} 
and just slightly higher than the experimental estimates.

 From this work we can conclude that realistic variational calculations for
medium--heavy, doubly--closed shell nuclei in {\it ls} coupling scheme 
with modern, sophisticated potentials are not only feasible, but have also 
reached the same degree of accuracy as in nuclear matter.

There are several natural extensions we envisage. For instance, the
 inclusion of three-body correlations, the study of N$\neq$Z closed
 shell nuclei and the development of the FHNC/SOC formalism for the 
{\it jj} coupling scheme. Work along these directions is in progress.

\section{Acknowledgements}

We want to thank Ingo Sick, Juerg Jourdan and Steven Pieper for providing 
us with the elastic electron scattering data, the Coulomb sum rule 
experimental analysis and the  $^{16}$O Monte Carlo results, respectively. 

\section{Appendix}

 In this Appendix we give the explicit expressions of the remaining 
diagrams contributing to the three--nucleon potential expectation 
value in FHNC/SOC.
\begin{eqnarray}
&\langle v^{2\pi}_{ijk}\rangle_{2.2} &= 6 A_{2\pi}
\int d^3r_1 \int d^3r_2  \int d^3r_3 
{ {f^{i_\sigma\tau}(r_{12})} \over {f^{c}(r_{12})} }
X^{l_\sigma}(r_{12}) 
{ {f^{j_\sigma\tau}(r_{13})} \over {f^{c}(r_{13})} }
X^{m_\sigma}(r_{13}) 
 \\ \nonumber & & 
\left[ 
R^{i_\sigma j_\sigma l_\sigma m_\sigma}
+3 A^{i_\sigma}\delta_{i_\sigma l_\sigma}A^{j_\sigma}\delta_{j_\sigma m_\sigma}
\right] \rho^c_3({\bf r}_1,{\bf r}_2,{\bf r}_3) ,
\label{TNI_22}
\end{eqnarray}
\begin{eqnarray}
&\langle v^{2\pi}_{ijk}\rangle_{2.3} = 12 A_{2 \pi}
&\int d^3r_1 \int d^3r_2  \int d^3r_3 
{ {f^{i_\sigma\tau}(r_{12})} \over {f^{c}(r_{12})} }
X^{l_\sigma}(r_{12}) 
{ {f^{j_\sigma\tau}(r_{23})} \over {f^{c}(r_{23})} }
X^{m_\sigma}(r_{13}) 
 \\ \nonumber & & 
\left[ 
L^{i_\sigma l_\sigma n_\sigma}
-K^{i_\sigma l_\sigma n_\sigma}A^{ n_\sigma}
\right]\xi^{j_\sigma m_\sigma n_\sigma}_{231}
\rho^c_3({\bf r}_1,{\bf r}_2,{\bf r}_3) ,
\label{TNI_23}
\end{eqnarray}
\begin{equation}
\langle v^{R}_{ijk}\rangle_{3.1} ={{1}\over{6}}
\int d^3r_1 \int d^3r_2  \int d^3r_3 
v^R_{123}
\rho^c_3({\bf r}_1,{\bf r}_2,{\bf r}_3) ,
\label{TNI_31}
\end{equation}
\begin{eqnarray}
&\langle v^{R}_{ijk}\rangle_{3.2} &={{1}\over{2}}
\int d^3r_1 \int d^3r_2  \int d^3r_3 
v^R_{123}
{ {f^i(r_{23})f^k(r_{23})} \over {[f^{c}(r_{23})]^2} }
 \\ \nonumber & & 
K^{ikm}A^m \left[ 
\rho^c_{3,dir}({\bf r}_1,{\bf r}_2,{\bf r}_3) \delta_{ik} + 
\rho^c_{3,exch}({\bf r}_1,{\bf r}_2,{\bf r}_3) \Delta^m \right] .
\label{TNI_32}
\end{eqnarray}
The $L^{ijk}$ and $R^{ijkl}$ matrices are given in Ref.\cite{pan79} and 
Ref.\cite{car83}, respectively. The central three--body density, 
$\rho^c_3({\bf r}_1,{\bf r}_2,{\bf r}_3)$, is written in 
superposition approximation as:
\begin{eqnarray}
&\rho^c_3({\bf r}_1,{\bf r}_2,{\bf r}_3) =
 &\sum_{xyzx'y'z'=d,e} 
 g^c_{xx'}({\bf r}_1,{\bf r}_2)
V^c_{x'y'}({\bf r}_2)
g^c_{y'y}({\bf r}_2,{\bf r}_3)
V^c_{yz}({\bf r}_3)
g^c_{zz'}({\bf r}_3,{\bf r}_1)
V^c_{z'x}({\bf r}_1) 
 \\ \nonumber & & 
 ~~~~~~~~~~-8 g^c_{cc}({\bf r}_1,{\bf r}_2)
V^c_{cc}({\bf r}_2)
g^c_{cc}({\bf r}_2,{\bf r}_3)
V^c_{cc}({\bf r}_3)
g^c_{cc}({\bf r}_3,{\bf r}_1)
V^c_{cc}({\bf r}_1) .
\label{rho3}
\end{eqnarray}

As already stated, only the $ee$ combination is not allowed at a given 
vertex. The  {\em exchange} and {\em direct} three--body densities, 
$\rho^c_{3,exch/dir}({\bf r}_1,{\bf r}_2,{\bf r}_3)$, are given by 
those parts of the full density where nucleons 2 and 3 belong or not 
to the same exchange loop.  


%
%
%
%
\newpage
\begin{table}
\caption{
TNI expectation values per nucleon, in MeV, in $^{16}$O and 
$^{40}$Ca for the Urbana V model. The $f_{1(6)}$ column gives the 
energies for the Jastrow (operatorial) correlation. The superscript  
$C(A)$ indicates the commutator (anticommutator) contribution. 
} 
\begin{tabular}{ccccc}
  & $f_1$(FHNC) & $f_1$(VMC) & $f_6$(SOC) & $f_6$(CMC) \\
\tableline
  & & $^{16}$O & & \\
 $\langle v^{2\pi ,C}_{ijk}\rangle$ & -0.17 & -0.16 &  -0.90 & -0.86 \\
 $\langle v^{2\pi ,A}_{ijk}\rangle$ &  0.74 &  0.70 &  -0.39 & -0.44 \\
 $\langle v^{R}_{ijk}\rangle$       &  1.33 &  1.28 &   1.65 &  1.57 \\
\tableline
  & & $^{40}$Ca & & \\
 $\langle v^{2\pi ,C}_{ijk}\rangle$ & -0.24 &       &  -1.50 &       \\
 $\langle v^{2\pi ,A}_{ijk}\rangle$ &  1.80 &       &  -0.26 &       \\
 $\langle v^{R}_{ijk}\rangle$       &  2.75 &       &   3.20 &       \\
\end{tabular}
\label{tab:3body}
\end{table}
%
%
%
\begin{table}
\caption{ 
FHNC/SOC and CMC energies in Mev/A for $^{16}$O with the 
A14+UVII interaction and the CMC single particle potential 
and nuclear matter correlation. 
$rms$ in fm. See text.
} 
\begin{tabular}{ccc}
   &  FHNC/SOC & CMC \\
 \tableline
  $\langle T \rangle - T_{cm}$        &  37.23 &  32.0 \\
  $\langle v_{6,ij}\rangle$           & -42.48 & -38.2 \\
  $\langle v_{7-8,ij}\rangle$         &  -0.85 &       \\
  $\langle v^{2\pi}_{ijk}\rangle$     &  -2.83 &       \\
  $\langle v^{R}_{ijk}\rangle$        &   1.90 &       \\
  $\langle v_{ijk}\rangle$            &  -0.93 &  -1.1 \\
  $\langle v_{Coul}\rangle$           &   1.00 &   0.9 \\
  $\Delta E$                          &   0.05 &       \\
  $\Delta E_{CMC}$                    &        &  -0.5 \\
  $E_{gs}$                            &  -5.97 &  -6.9 \\
  $rms$                               &   2.44 &   2.43\\
\end{tabular}
\label{tab:cmc}
\end{table}
\begin{table}
\caption{ 
Nuclear matter energies per nucleon 
for  A18+UIX (A18), A8'+UIX (A8') and A14+UVII(A14).
The $\Delta$E/A columns list the spin--orbit correlation
and the $p>$8 potential components corrections.
Densities in fm$^{-3}$, energies in MeV. 
} 
\begin{tabular}{cccccc}
 $\rho_{nm}$  &  E/A(A18)  & E/A (A8')   & $\Delta$E/A (A8') 
               & E/A(A14)  & $\Delta$E/A (A14)  \\
\tableline
 0.04  & -4.11 & -5.50 & -0.30 & -5.25 & -0.80  \\
 0.08  & -7.46 & -8.45 & -0.82 & -8.43 & -0.38  \\
 0.12  & -9.42 &-10.31 & -1.48 &-10.63 &  0.00  \\
 0.16  &-10.05 &-10.87 & -2.16 &-11.99 &  0.61  \\
 0.20  & -8.74 &-10.06 & -3.02 &-12.37 &  0.27  \\
 0.24  & -5.66 & -7.75 & -3.96 &-11.74 &  0.50  \\
\end{tabular}
\label{tab:nm}
\end{table}
%
%
%
\begin{table}
\caption{ 
Energies in MeV/A for $^{16}$O and $^{40}$Ca obtained    
with A8'+UIX+$f_{6,nm}$ and with truncated A14+UVII+$f_{6,nm}$.
In the upper part of the table we show the values
the minimization parameters.
$b_{HO}$, $a_0$, $R_0$, are expressed in fm, $\rho_{nm}$ in fm$^{-3}$
all the other quantitites in MeV/A.
The energy experimental values are -7.97 MeV/A for $^{16}$O and 
-8.55 MeV/A for $^{40}$Ca. The experimental rms radii are 2.73 and 3.48
fm for  $^{16}$O and  $^{40}$Ca, respectively.  
} 
\begin{tabular}{ccccccc}
  & $^{16}$O(HO)$_{A8'}$  & $^{16}$O(WS)$_{A8'}$  & $^{16}$O(WS)$_{A14}$  
  & $^{40}$Ca(HO)$_{A8'}$ & $^{40}$Ca(WS)$_{A8'}$ & $^{40}$Ca(WS)$_{A14}$  \\
\tableline
 $b_{HO}$                        &   2.00 &        & 
                                 &   2.10 &        & \\
 $V_0$                           &        & -42.00 & -36.00  
                                 &        & -50.00 & -41.50 \\
 $R_0$                           &        &   3.60 & 3.80
                                 &        &   5.30 & 5.00\\
 $a_0$                           &        &   0.55 & 0.55
                                 &        &   0.53 & 0.55 \\
 $\rho_{nm}$                     &   0.09 &   0.09 &  0.09
                                 &   0.09 &   0.09 &  0.09 \\
\tableline
 $\langle T \rangle$             &  22.57 &  27.34 &  25.91
                                 &  30.02 &  30.58 &  30.90\\
%
%
 $\langle T \rangle_{IPM}$       &  11.64 &  13.82 &  12.20
                                 &  14.09 &  14.15 &  14.10\\
%
%
 $\langle v_{8,ij}\rangle$       & -27.49 & -32.48 & -31.05
                                 & -38.03 & -38.68 & -39.08 \\
 $\langle v^{2\pi}_{ijk}\rangle$ &  -1.15 &  -1.66 & -1.35
                                 &  -1.94 &  -1.94 & -1.74 \\
 $\langle v^{R}_{ijk}\rangle$    &   1.26 &   1.98 & 1.56
                                 &   2.46 &   2.37 & 2.11\\
 $E_{cm}$                        &   0.48 &   0.58 & 0.52
                                 &   0.18 &   0.20 & 0.19 \\
 $\langle v_{Coul}\rangle$       &   0.78 &   0.86 & 0.84
                                 &   1.85 &   1.85 & 1.87 \\
 $\Delta E$                      &  -0.66 &  -0.94 & -0.50
                                 &  -0.95 &  -0.96 & -0.70 \\
 $E_{gs}$                        &  -5.18 &  -5.48 & -5.11
                                 &  -6.77 &  -6.97 & -6.50 \\
 $rms$                           &   3.03 &   2.83 & 2.93
                                 &   3.65 &   3.66 & 3.66 \\
\end{tabular}
\label{tab:hows}
\end{table}
\begin{table}
\caption{ 
Energies in MeV/A for $^{16}$O and $^{40}$Ca obtained    
with A8'+UIX+$f_{6,nm}$
and with WS mean field potentials fixed to reproduce
the empirical charge densities.
$R_0$ and $a_0$ are expressed in fm, 
$\rho_{nm}$ in fm$^{-3}$, and the other quantities in MeV/A.
} 
\begin{tabular}{ccc}
  & $^{16}$O(WS)$_{A8'}$  & $^{40}$Ca(WS)$_{A8'}$  \\
\tableline
 $V_0$                           & -53.00 & -50.00 \\
 $R_0$                           &   3.45 &   4.60 \\
 $a_0$                           &   0.7  &   0.5  \\
 $\rho_{nm}$                     &   0.09 &  0.09 \\
\tableline
 $\langle T \rangle$             &  32.64 &  38.15 \\
%
%
 $\langle T \rangle_{IPM}$       &  15.35 &  16.82 \\
%
%
 $\langle v_{8,ij}\rangle$       & -37.79 & -46.34 \\
 $\langle v^{2\pi}_{ijk}\rangle$ &  -2.36 &  -2.98 \\
 $\langle v^{R}_{ijk}\rangle$    &   3.00 &   3.94 \\
 $E_{cm}$                        &   0.64 &   0.23 \\
 $\langle v_{Coul}\rangle$       &   0.94 &   2.10 \\
 $\Delta E$                      &  -1.21 &  -1.28 \\
 $E_{gs}$                        &  -5.41 &  -6.64 \\
 $rms$                           &   2.67 &   3.39  \\
\end{tabular}
\label{tab:wsfit}
\end{table}
%
%
%
%
\newpage
\begin{figure}
\caption{
Cluster diagrams considered for the three-body force expectation value. 
The $2.1-3$ diagrams are
related to the  $\langle v^{2\pi}_{ijk}\rangle$ part of the force and
the $3.1-2$ diagrams are related to  $\langle v^R_{ijk} \rangle$.
The points denote the particle coordinates. The dashed, wavy 
and double--wavy lines
denote generalized scalar, operator and single--operator ring 
correlation bonds, respectively. 
See Ref. \protect\cite{car83} for more details.
}
\label{fig:diag}
\end{figure}

\begin{figure}
\caption{
FHNC/SOC charge densities related to the results of
Tab. \protect\ref{tab:hows} (thin full lines) compared with the IPM densities
(dashed lines) and with the empirical ones (thick full lines).
}
\label{fig:dens6}
\end{figure}
\begin{figure}
\caption{
The same as Figure \protect\ref{fig:dens6} for the calculation of 
Tab. \protect\ref{tab:wsfit}
}
\label{fig:dens2}
\end{figure}
\begin{figure}
\caption{
Electron scattering elastic cross sections for $^{16}$O (left) and 
$^{40}$Ca (right). 
The full lines have been produced by the FHNC/SOC densities of
Figure \protect\ref{fig:dens2}. The other two lines by the WS
FHNC/SOC densities of Figure \protect\ref{fig:dens6}. The dashed lines
correspond to the A8' densities while the dotted lines to the A14 ones. 
} 
\label{fig:cross}
\end{figure}
\begin{figure}
\caption{
Two--nucleon (upper panels) and proton--proton (lower panels)
distribution functions for the A8'+UIX interaction. 
The dashed lines give the IPM results, 
the solid lines the FHNC/SOC ones. 
}
\label{fig:tbd}
\end{figure}
\begin{figure}
\caption{
Density (upper panel) and longitudinal (lower panel) structure 
functions for $^{16}$O (left) for the A8'+UIX interaction. 
The IPM (dashed lines), FHNC/SOC (solid lines), non--normalized FHNC/SOC 
(triangles) and Jastrow (stars) results are given.
} 
\label{fig:sfsl}
\end{figure}
\begin{figure}
\caption{
Coulomb sums for $^{16}$O, $^{40}$Ca and nuclear matter ({\it n.m.}) 
compared with the experimental results from $^{12}$C, $^{40}$Ca  
and $^{56}$Fe.
} 
\label{fig:csr}
\end{figure}
\begin{figure}
\caption{
Isovector spin longitudinal (solid) and transverse (dashed) 
structure functions for the  $^{16}$O and $^{40}$Ca nuclei.
For $^{16}$O we show also the Jastrow results (dotted line). 
The lowest panel gives the
ratios of the isovector spin longitudinal and transverse 
structure functions for both nuclei. 
} 
\label{fig:spinsf}
\end{figure}

\begin{references}

\bibitem{che86} C. R. Chen, G. L. Payne, J. L. Friar and B. F. Gibson, 
Phys. Rev. C {\bf 33}, 1740 (1986); 
 A. Stadler, W. Gl\"ockle and P. U. Sauer, Phys. Rev. C {\bf 44}, 2319 (1991).
\bibitem{kie93} A. Kievsky, M. Viviani and S. Rosati, 
 Nucl. Phys. A {\bf 551}, 241 (1993).
\bibitem{pud97} B. S. Pudliner, V. R. Pandharipande, J. Carlson, S. C. Pieper 
 and R. B. Wiringa, Phys. Rev. C {\bf 56}, 1720 (1997).
\bibitem{wir88} R. B. Wiringa, V. Ficks and A. Fabrocini, 
 Phys. Rev. C {\bf 38}, 1010 (1988).
\bibitem{akm98} A. Akmal, V. R. Pandharipande and  D. G. Ravenhall, 
 Phys. Rev. C {\bf 58}, 1140 (1998).
\bibitem{fri81} B. A. Friedman and Pandharipande,
 Nucl. Phys. A {\bf 361}, 502 (1981).
\bibitem{akm97} A. Akmal and V. R. Pandharipande, 
 Phys. Rev. C {\bf 55}, 2261 (1997).
\bibitem{fab89} A. Fabrocini and S. Fantoni, 
 Nucl. Phys. A {\bf 503}, 375 (1989).
\bibitem{fab97} A. Fabrocini, 
 Phys. Rev. C {\bf 55}, 330 (1997).
\bibitem{ben92} O. Benhar, A. Fabrocini and S. Fantoni, 
 Nucl. Phys. A {\bf 550}, 201 (1992).
\bibitem{arn92} R. A. Arndt, L. D. Roper, R. L. Workman and 
 M. W. McNaughton, Phys. Rev. D {\bf 45}, 3995 (1992).
\bibitem{sto93} V. G. J. Stoks, R. A. M. Klomp, M. C. M. Rentmeester and 
 J. J. DeSwart, Phys. Rev. C {\bf 48}, 792 (1993).
\bibitem{sto94} V. G. J. Stoks, R. A. M. Klomp, C. P. F. Terheggen  and 
 J. J. DeSwart, Phys. Rev. C {\bf 49}, 2950 (1994).
\bibitem{wir95} R. B. Wiringa, V. G. J. Stoks and R. Schiavilla, 
 Phys. Rev. C {\bf 51}, 38 (1995).
\bibitem{mac96} R. Machleidt, F. Sammarruca and Y. Song,  
 Phys. Rev. C {\bf 53}, R1483 (1996).
\bibitem{wir91} R.B. Wiringa, Phys. Rev. C {\bf 43}, 1585 (1991).
\bibitem{pie92} S. C. Pieper, R. B. Wiringa and V. R. Pandharipande, 
 Phys. Rev. C {\bf 46}, 1741 (1992).
\bibitem{ros82} S. Rosati, in {\em From nuclei to particles}, 
 Proc. Int. School E. Fermi, course LXXIX, 
ed. A. Molinari (North Holland, Amsterdam, 1982).
\bibitem{pan79} V. R. Pandharipande and R. B. Wiringa, 
 Rev. Mod. Phys.  {\bf 51}, 821 (1979).
\bibitem{wir80}  R. B. Wiringa, 
 Nucl. Phys. A {\bf 338}, 57 (1980).
\bibitem{co92} G. Co', A. Fabrocini, S. Fantoni and I. E. Lagaris, 
 Nucl. Phys. A {\bf 549}, 439 (1992).
\bibitem{co94} G. Co', A. Fabrocini and S. Fantoni, 
 Nucl. Phys. A {\bf 568}, 73 (1994).
\bibitem{ari96} F. Arias de Saavedra, G. Co', A. Fabrocini and S. Fantoni, 
 Nucl. Phys. A {\bf 605}, 359 (1996).
\bibitem{fab98} A. Fabrocini,  F. Arias de Saavedra, G. Co' and 
 P. Folgarait, Phys. Rev. C {\bf 57}, 1668 (1998).
\bibitem{wir84} R. B. Wiringa, R. A. Smith and T. L. Ainsworth, 
 Phys. Rev. C {\bf 29}, 1207 (1984).
\bibitem{sch86} R. Schiavilla, V. R. Pandharipande and  R. B. Wiringa, 
 Nucl. Phys. A {\bf 449}, 219 (1986).
\bibitem{fuj57} J. Fujita and H. Miyazawa, 
 Progr. Ther. Phys. {\bf 17}, 360 (1957).
\bibitem{car83} J. Carlson, V. R. Pandharipande and R. B. Wiringa,  
 Nucl. Phys. A {\bf 401}, 59 (1983).
\bibitem{lag81} I. E. Lagaris and V. R. Pandharipande, 
 Nucl. Phys. A {\bf 359}, 331 (1981);
 Nucl. Phys. A {\bf 359}, 349 (1981).
\bibitem{lag80} I. E. Lagaris and V. R. Pandharipande, 
 Nucl. Phys. A {\bf 334}, 217 (1980).
\bibitem{pie99} S. C. Pieper, private communication.
\bibitem{hei99} J. H. Heisenberg and B. Mihaila, 
 Phys. Rev. C {\bf 59}, 1440 (1999).
\bibitem{dej87} C.W. De Jager and C. De Vries,
At. Data and Nucl. Data Tables, {\bf 36}, 495 (1987).
\bibitem{hoe76} G. Hoehler, E. Pietarinen, I. Sabba-Stefabescu,
  F. Borkowski, G.G. Simon, V.H. Walther and R.D. Wendling,
  Nucl. Phys. B {\bf 114},505 (1976).
\bibitem{ann95} R. Anni, G. Co' and P. Pellegrino, Nucl. Phys. A 
{\bf  584}, 35 (1995); R. Anni and  G. Co' Nucl. Phys. A {\bf  588},
463 (1995). 
\bibitem{sic70} I. Sick and J.S. McCarthy, Nucl. Phys. A {\bf 150},
                631 (1970);
                B.B.P. Sinha, G.A. Peterson, R.R. Whitney I. Sick and
                J.S. McCarthy, Phys. Rev. C {\bf 7}, 1930 (1973);
                I. Sick, J. Bellicard, J. Cavedon, B. Frois, M Huet,
                P. Leconte, P. Ho and S. Platchkov, Phys. Lett. B 
                {\bf 88}, 245 (1979);
                 J. Cavedon, Th\`ese de doctorat d'Etat, Paris 1980,
                 unpublished; I. Sick private communication.
\bibitem{ari97} F. Arias de Saavedra, G. Co' and M.M. Renis,
 Phys. Rev. C {\bf 55}, 673 (1997).
\bibitem{jou96} J. Jourdan, 
 Nucl. Phys. A {\bf 603}, 117 (1996); private communication.
\bibitem{sch87} R. Schiavilla, D. S. Lewart, V. R. Pandharipande, 
S. C. Pieper  and  R. B. Wiringa, 
 Nucl. Phys. A {\bf 473}, 267 (1987).
\bibitem{mcc92} J. M. McClelland {\em et al.}, 
 Phys. Rev. Lett. {\bf 69}, 582 (1992).
\bibitem{alb82} W. M. Alberico, M. Ericson and A. Molinari, 
 Nucl. Phys. A {\bf 379}, 429 (1982).
\bibitem{ich89} M. Ichimura, K. Kawahigashi, 
 T. S. Jorgensen and C. Gaarde,
 Phys. Rev. C {\bf 39}, 1446 (1989).
\bibitem{fab94} A. Fabrocini, 
 Phys. Lett. B {\bf 322}, 171 (1994).
\bibitem{pan94} V. R. Pandharipande, J. Carlson, S. C. Pieper, 
 R. B. Wiringa and R. Schiavilla, 
 Phys. Rev. C {\bf 49}, 789 (1994).
\bibitem{fan83} S. Fantoni, B. L. Friman  and V. R. Pandharipande, 
 Nucl. Phys. A {\bf 399}, 57 (1983).
\bibitem{fab93} A. Fabrocini and S. Fantoni, 
 Phys. Lett. B {\bf 298}, 263 (1993).

\end{references}
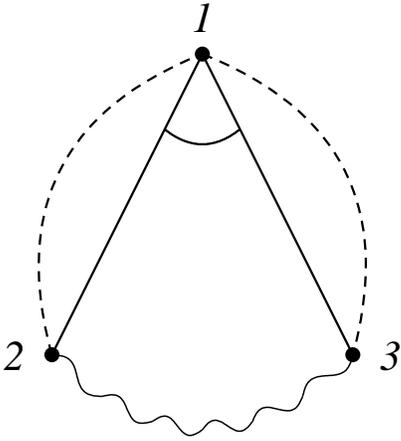
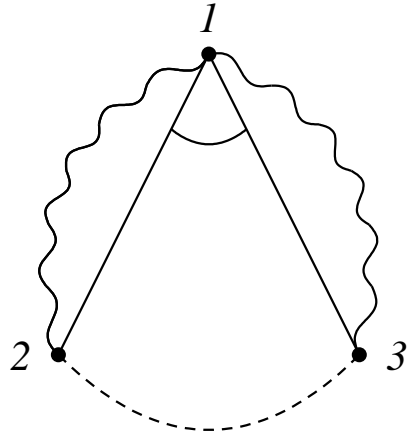
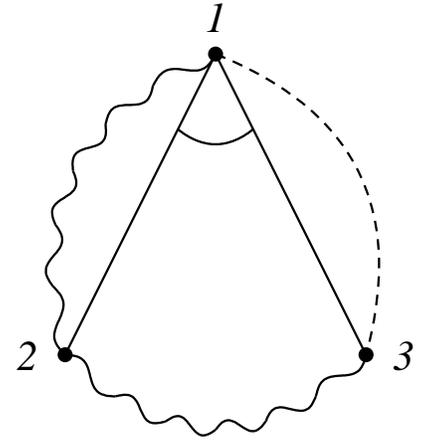
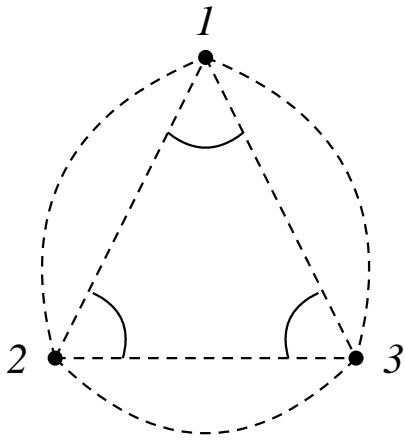
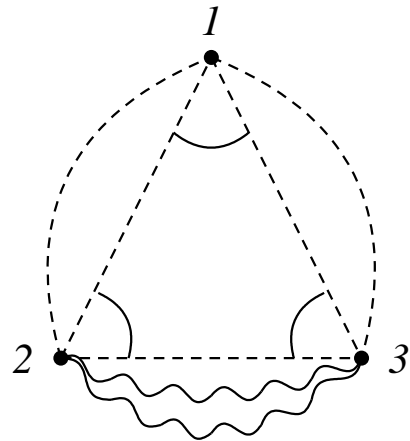
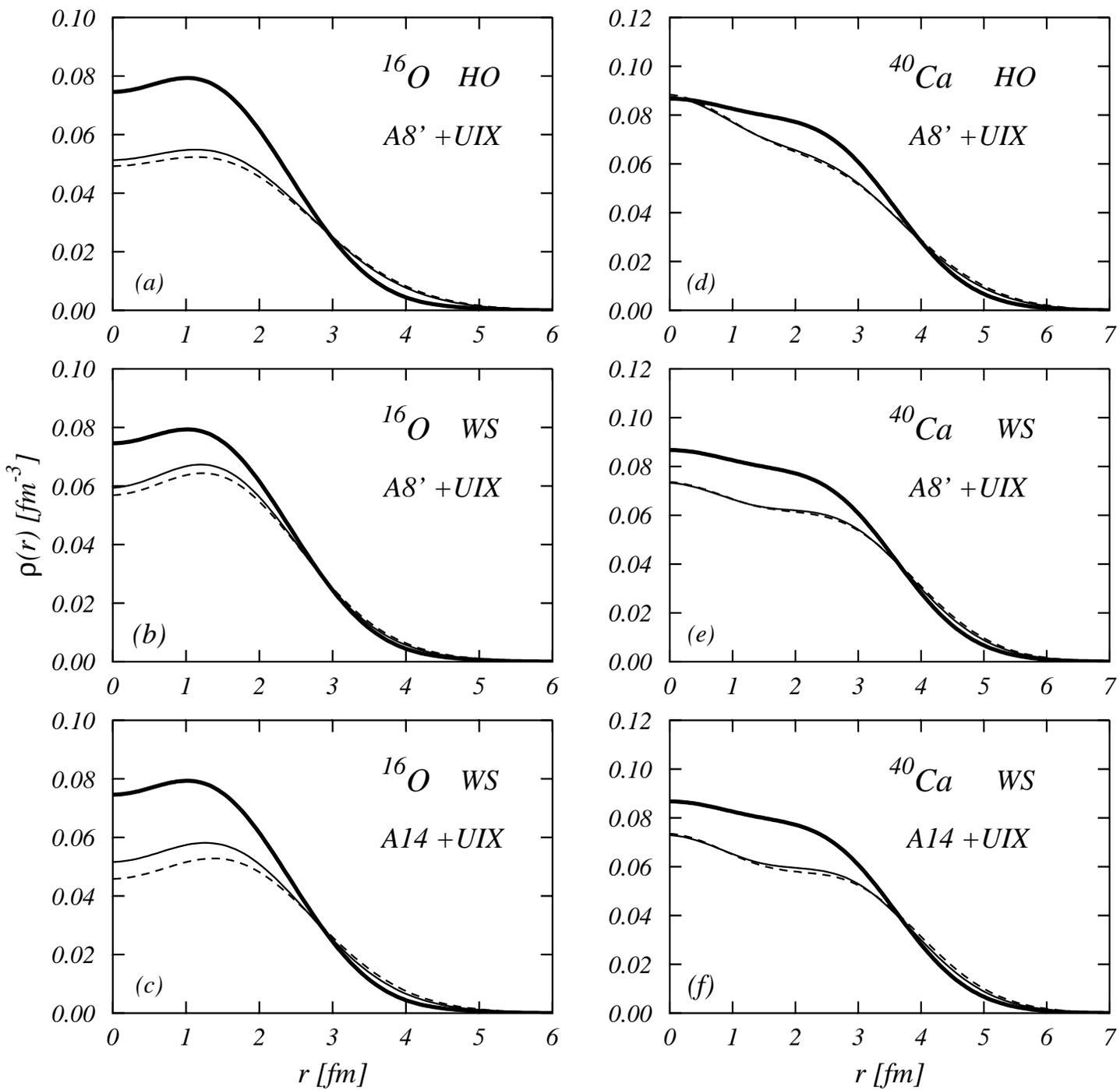
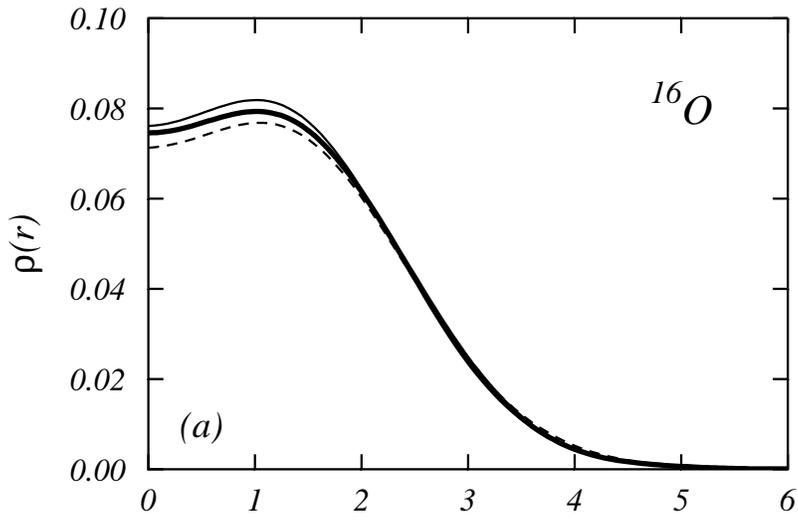
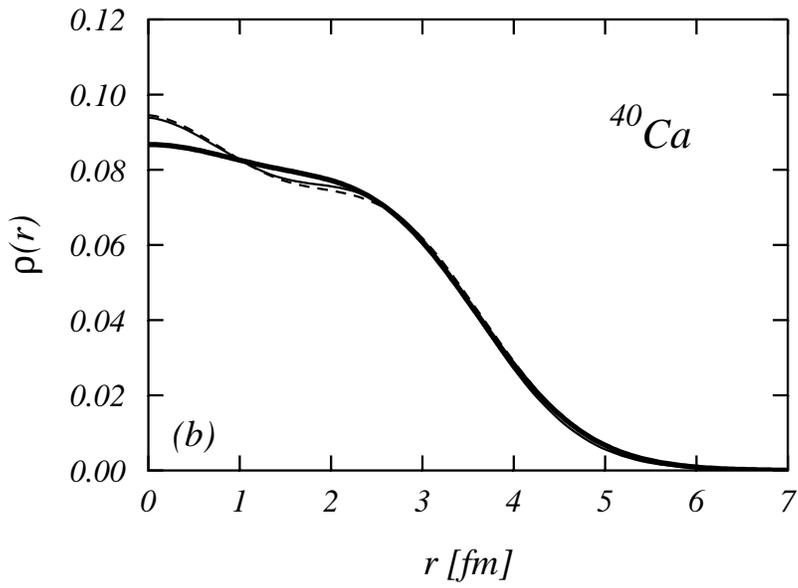
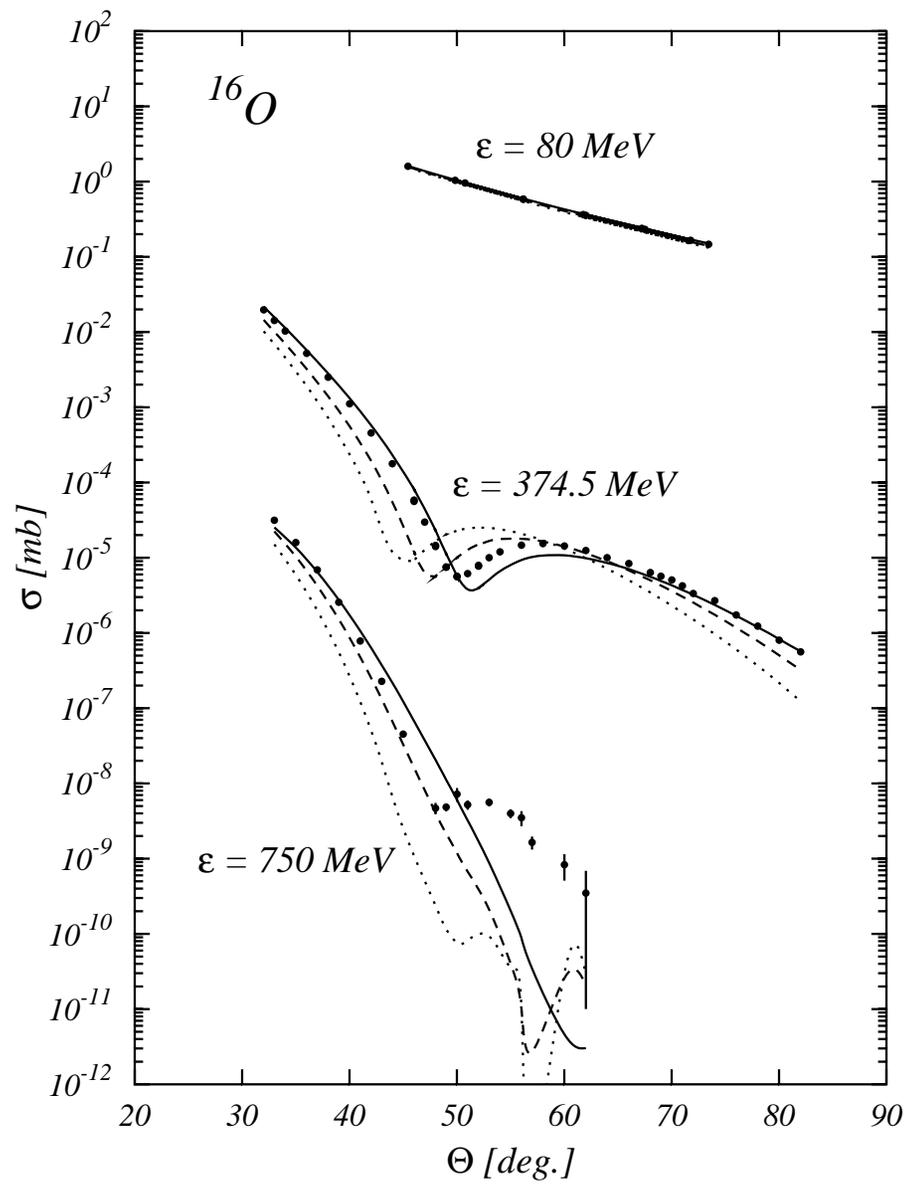
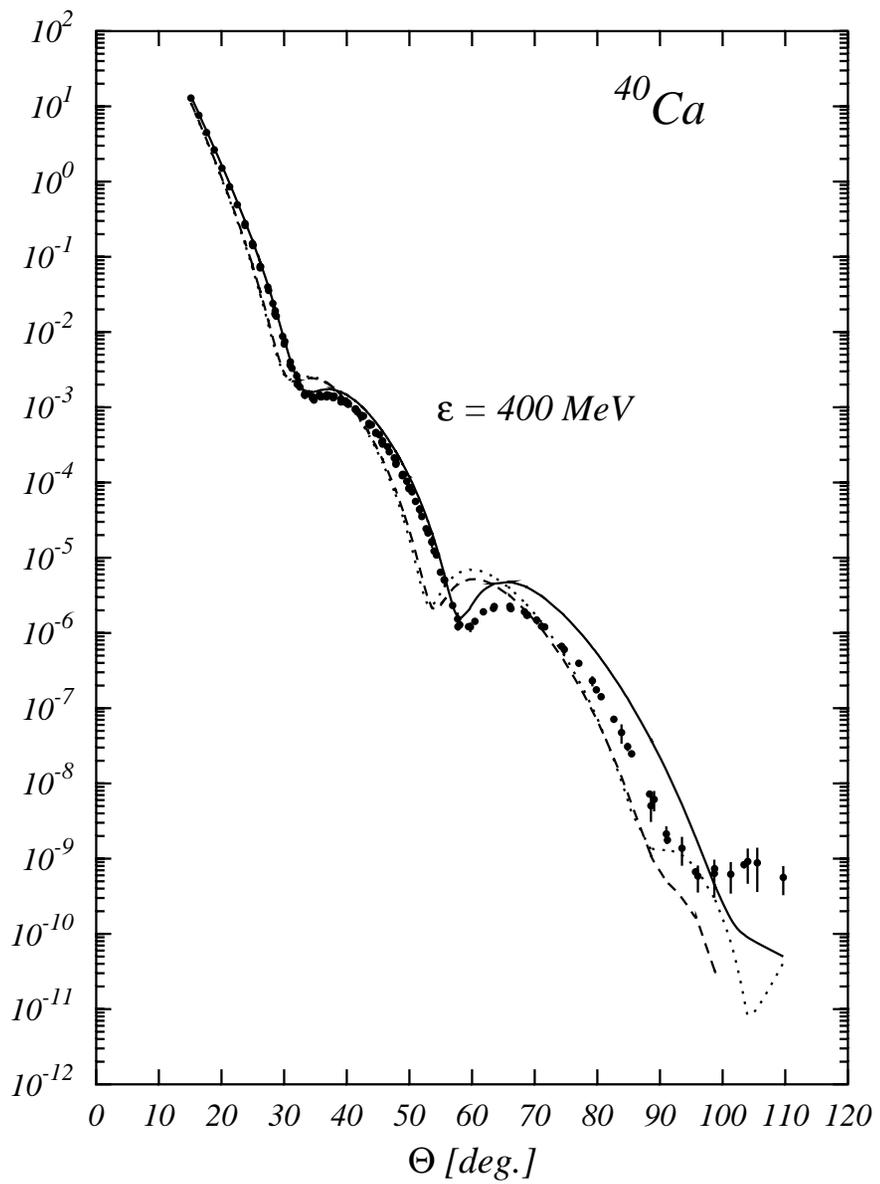
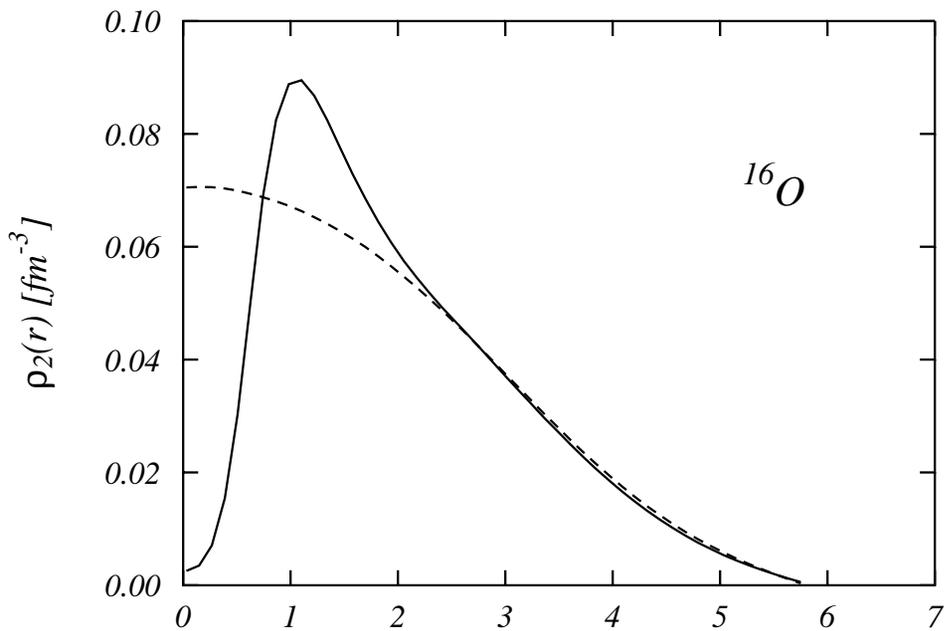
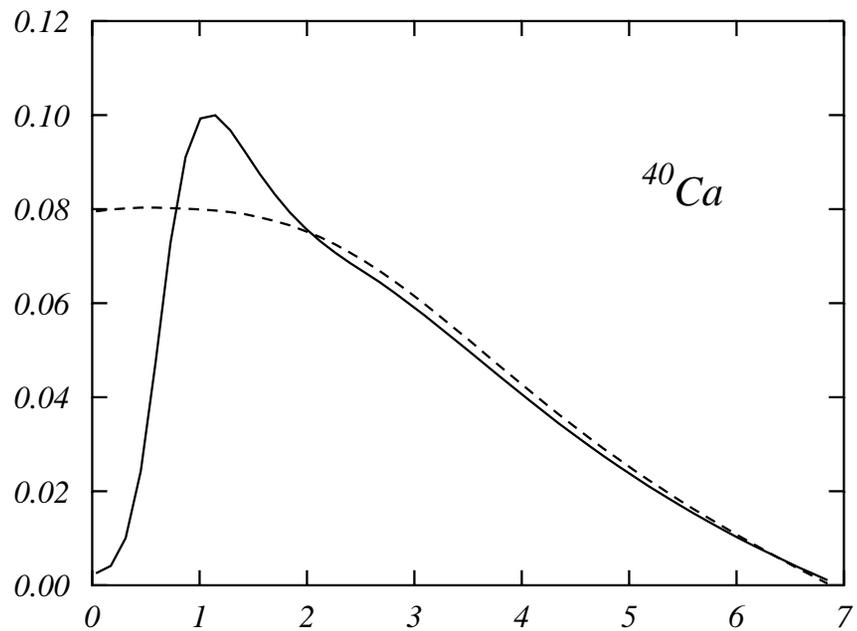
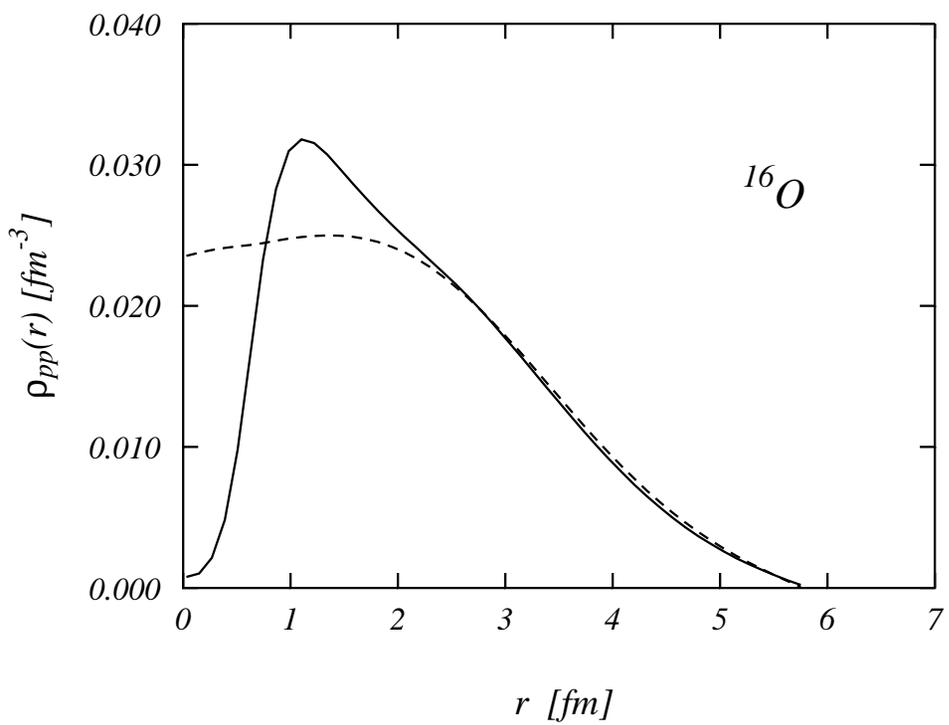
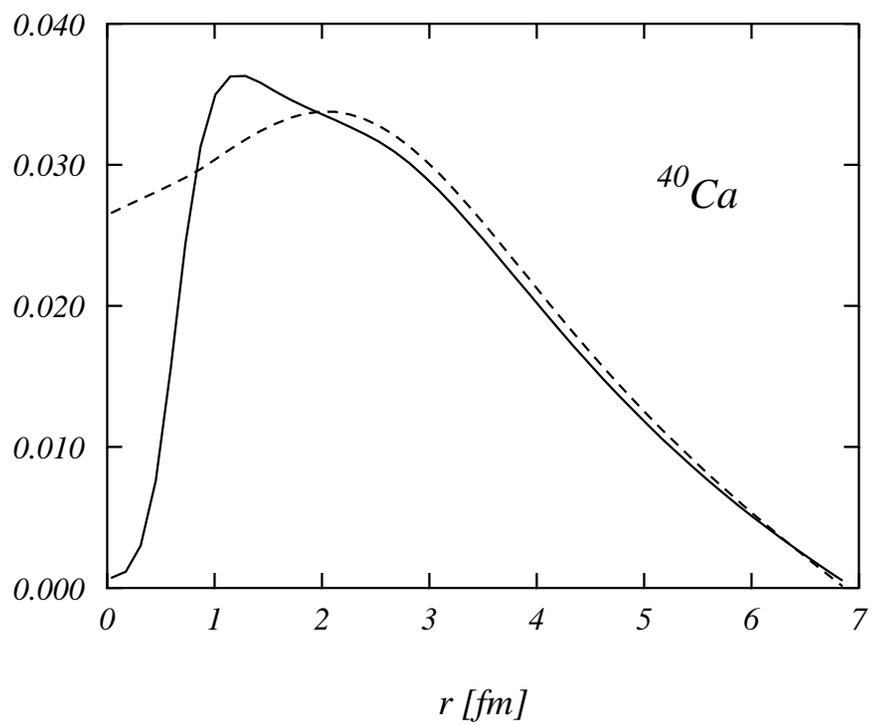
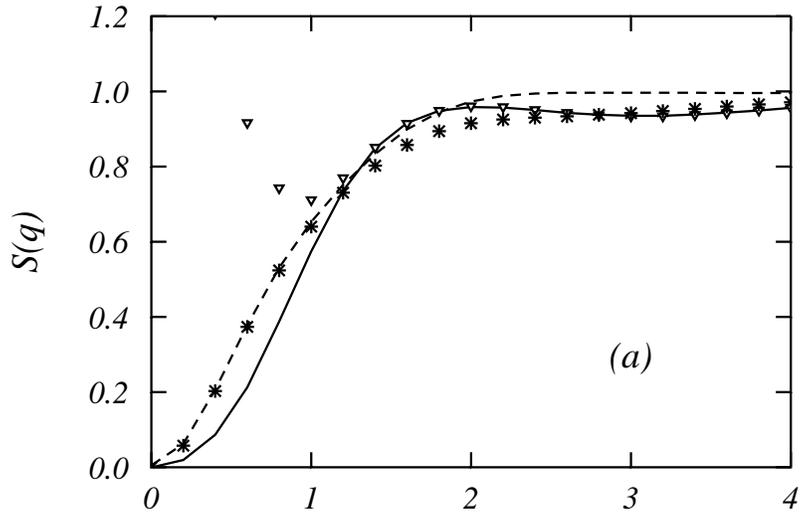
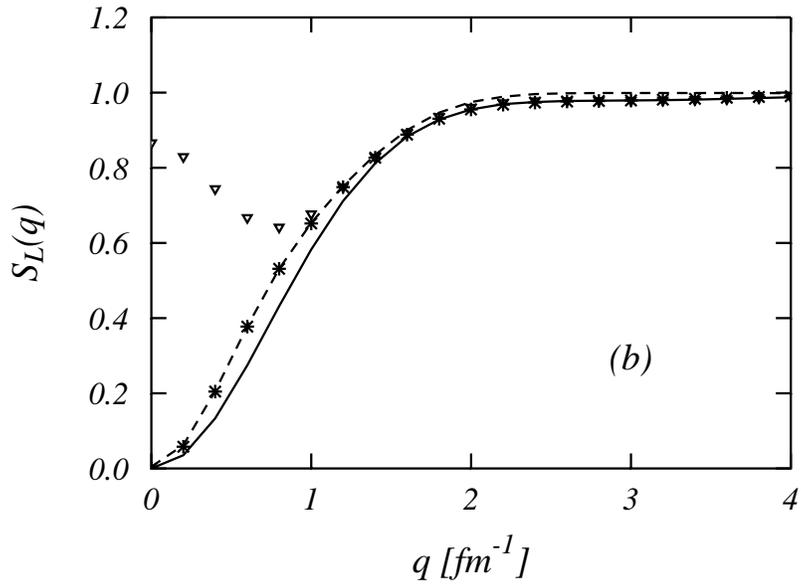
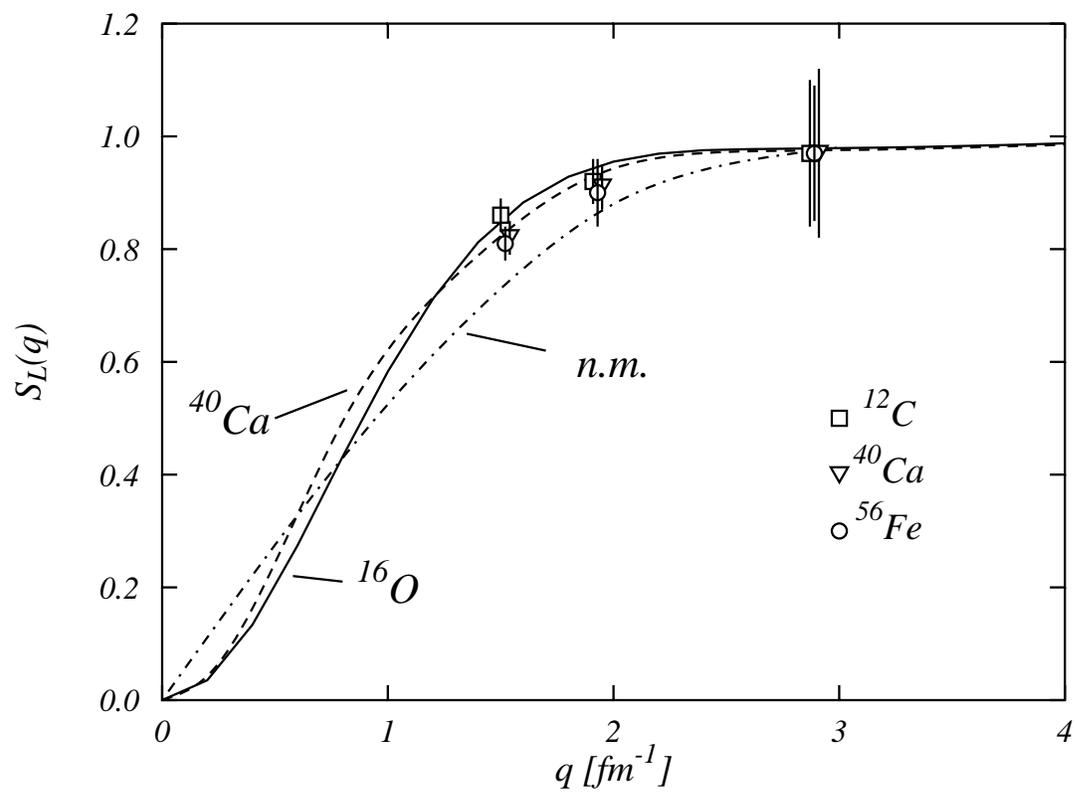
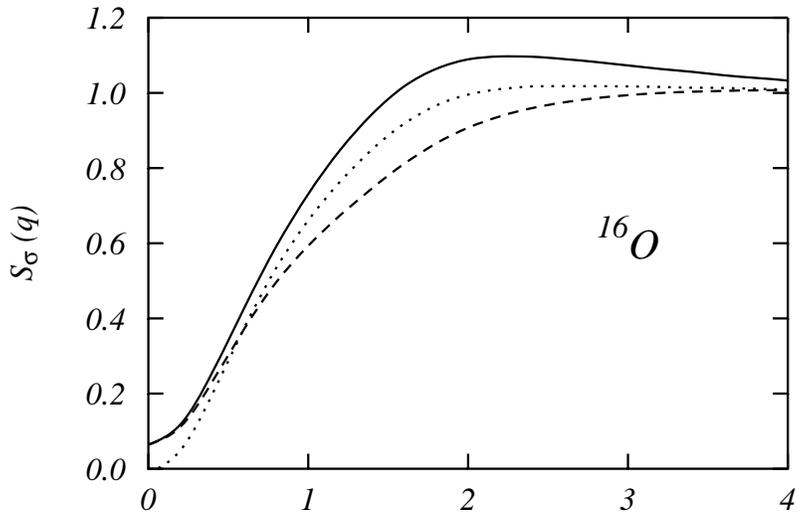
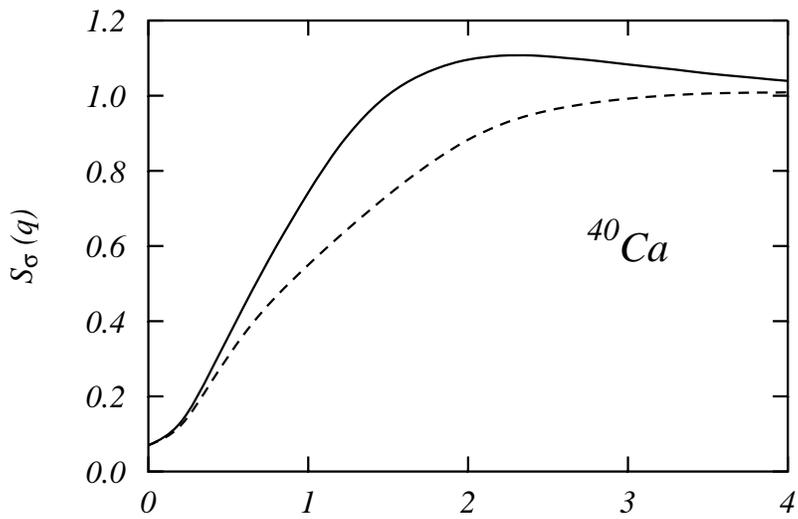
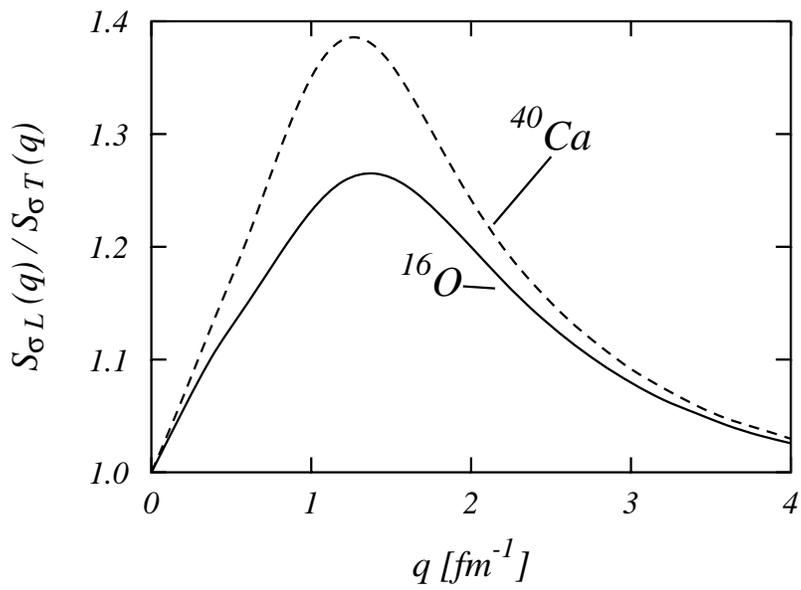
\end{document}